\documentclass[12pt,preprint]{aastex}

\input epsf
\usepackage{epsfig}
\usepackage{rotating}

\slugcomment{Submitted to Astrophysical Journal}

\shorttitle{Galaxy Clusters in the Swift/BAT era}
\shortauthors{Ajello et al.}

\begin{document}


\title{Galaxy Clusters in the {\it Swift}/BAT era II: 10 more
Clusters detected above 15\,keV.}

\author{M. Ajello\altaffilmark{1}, P. Rebusco\altaffilmark{2},
N. Cappelluti\altaffilmark{3,4},  O. Reimer\altaffilmark{1,5},
H. B\"ohringer\altaffilmark{3}, V. La Parola\altaffilmark{6} and
G. Cusumano\altaffilmark{6}.
}

\email{majello@slac.stanford.edu}
%%%%%%%%%%%%%%%%%%%%%%%%%%%%%%%
\altaffiltext{1}{SLAC National Laboratory and Kavli Institute
for Particle Astrophysics and Cosmology, 2575 Sand Hill Road, Menlo Park,
CA 94025, USA}

\altaffiltext{2}{Kavli Institute for Astrophysics and Space Research, MIT, Cambridge, MA 02139, USA}

\altaffiltext{3}{Max Planck Institut f\"{u}r Extraterrestrische Physik, P.O. Box 1603, 85740, Garching, Germany}

\altaffiltext{4}{University of Maryland, Baltimore County, 1000 Hilltop Circle, Baltimore, MD 
21250}

\altaffiltext{5}{Institut f\"ur Astro- und Teilchenphysik  
Leopold-Franzens-Universit\"at Innsbruck Technikerstr. 25/8, 6020 Innsbruck,
Austria }
\altaffiltext{6}{
INAF, Istituto di Astrofisica Spaziale e Fisica Cosmica di Palermo, via U. La Malfa 153, 90146 Palermo, Italy}

%\altaffiltext{6}{INAF-OAB, via E. Bianchi 46, Merate (LC) 23807,  Italy}
%%%%%%%%%%%%%%%%%%%%%%%%%%%%%%%%%%%%%%%%%%%%%
%
% History of changes:
%                      01-10-2009:pr edits cluster description
%                      01-11-2009:ma edits first part
%                      02-12-2009:ma+pr put up the first draft
%
%
%                      22-02-2010:ma finished (and re-did) the analysis
%                                 of all 8 clusters.
%         
%                      08-04-2010: ma finished to implement all comments
% 
%%%%%%%%%%%%%%%%%%%%%%%%%%%%%%%%%%%%%%%%%%%%%%%%%%%%%%%%%%%%%%%%%%%%%%%%%

%%%%%%%%%%%%%%%%%%%%%%%%%%%%%%%%%%%%%%%%%%%%%%%%%%%%%%%%%%%%%%%%%%%%%%%%%%%
%
%                   ABSTRACT
%
%%%%%%%%%%%%%%%%%%%%%%%%%%%%%%%%%%%%%%%%%%%%%%%%%%%%%%%%%%%%%%%%%%%%%%%%%%%

\begin{abstract}
We report  on the discovery of 10 additional  galaxy clusters detected
in the ongoing {\it Swift}/BAT all-sky survey. Among the newly BAT-discovered
clusters there are: Bullet, Abell 85, Norma, and PKS 0745-19.
Norma is the only cluster, among those presented here, which is resolved by BAT.
For all the clusters we perform a detailed spectral analysis
using XMM-Newton and {\it Swift}/BAT data to investigate the presence of  
a hard (non-thermal) X-ray excess. We find that in most cases 
the clusters' emission in the 0.3--200\,keV band can be explained
by a multi-temperature thermal model confirming
our previous results.  For two clusters (Bullet  and 
Abell 3667) we find evidence for the presence of 
a hard X-ray excess. In the case of the Bullet cluster,
our analysis confirms the presence of a non-thermal, power-law like,
component with a 20--100\,keV flux of 3.4$\times10^{-12}$\,erg cm$^{-2}$ s$^{-1}$
as detected in previous studies. For Abell 3667 the excess emission
can be successfully modeled as a hot component (kT=$\sim$13\,keV).
We thus conclude that the hard X-ray
emission from galaxy clusters (except the Bullet)
has most likely 
thermal origin. 
%This is in agreement with recent studies that 
%ound out that the pressure of all non-thermal phenomena to the 
%intracluster medium
%s likely $<$10\,\%.
\end{abstract}


%%%%%%%%%%%%%%%%%%%%%%%%%%%%%%%%%%%%%%%%%%%%%%%%%%%%%%%%%%%%%%%%%%%%%%%%%%%
%
%                   INTRODUCTION
%
%%%%%%%%%%%%%%%%%%%%%%%%%%%%%%%%%%%%%%%%%%%%%%%%%%%%%%%%%%%%%%%%%%%%%%%%%%%

\section{Introduction}
\label{intro}

The study of clusters of galaxies at X-ray energies
is key to
understand the mechanisms that heat the intra-cluster  medium (ICM) 
and to measure the pressure due to cosmic rays (CRs), magnetic fields and 
turbulence. In particular shock heating can be influenced by CRs 
if a significant part of the shock energy is transferred to charged particles.
Indeed, large-scale shocks that form during the process of cluster formation 
are believed to be efficient particle accelerators 
\citep[e.g.][]{sar99,ryu03}. Thus the pressure support of CRs to the
ICM might be relevant.

Not surprisingly, 
the role of CRs in the formation and evolution of clusters of galaxies
has been much debated. 
\cite{chu08} suggest that in massive galaxy clusters
hydrostatic equilibrium is satisfied reasonably well, as long as the 
source has not experienced a recent major merger.
On the other hand, other studies \citep[e.g.][]{mir95,nag07} showed that 
%However, in  relaxed and non-relaxed clusters in particular,
the non-thermal pressure due to CRs, 
magnetic fields and micro-turbulence can affect  the mass estimates 
based on hydrostatic equilibrium.% \citep[e.g.][]{mir95,nag07}, leading to a  higher baryonic to total mass ratio.
Knowing the importance of CRs, the mechanisms that heat the ICM and the 
frequency at which it is shocked, is crucial for the upcoming X-ray and 
Sunyaev-Zeldovich effect cluster surveys \citep[see][]{ando08}.

The detections of extended synchrotron radio emissions 
\citep[e.g.][]{wil70,har78,gio93,gio00,kem01,thierbach03} 
represent the main evidence that a  population of non-thermal
relativistic electrons exists in the ICM.
These very same electrons can produce X-rays via Inverse Compton (IC)
scattering off cosmic microwave background (CMB) photons 
\citep[e.g.][]{rep79,sar99}, or via 
non-thermal bremsstrahlung \citep[e.g.][]{sar99,sar00} 
or synchrotron radiation \citep[][]{tim04,ino05}.
Detecting this non-thermal radiation is difficult because of the 
bright and dominant ICM thermal emission. Studying clusters
above 15\,keV, where the intensity of the thermal component decreases
quickly, might prove to be an effective probe of the non-thermal
emission processes. Indeed, in the past, 
the detection of non-thermal emission in the hard X-ray spectra
of a few galaxy clusters has been reported \citep[see e.g.][for a complete review]{kaastra08}. However its actual presence and origin remain controversial 
\citep[e.g.][]{rephaeli87,rephaeli99,rephaeli02,
ros04,san05,renaud06,fus07,lutovinov08,molendi08}.

In a first paper \citep{ajello09a}, we reported about the detailed
analysis of 10 galaxy clusters serendipitously detected in the {\it Swift}/BAT 
all-sky survey above 15\,keV. In that study we concluded that 
there were no significant evidences for the existence of
hard X-ray excesses detected in the spectra of clusters
above the BAT sensitivity.
In this paper we report the analysis of 10 additional clusters that
have been recently  detected, thanks to the deeper exposure,
in the ongoing BAT survey.
We combine BAT and XMM-Newton data to find the best spectral fit. 
Assuming that there is a  non-thermal emission  due to 
IC scattering on CMB photons, we estimate the upper limit
of its flux in the 50-100 kev band. This information allows us 
to  estimate the intensity of the magnetic fields in these galaxy clusters. 

Throughout this paper 
we adopt a Hubble constant of $H_0$ = 70\,km\,s$^{-1}$\,Mpc$^{-1}$, 
$\Omega_M$ = 0.3 and $\Omega_\Lambda$ = 0.7.
Unless otherwise stated errors are quoted at the 90\,\% confidence
level (CL) for one interesting parameter and solar abundances
are determined using the meteoritic values provided in \cite{anders89}.

%%%%%%%%%%%%%%%%%%%%%%%%%%%%%%%%%%%%%%%%%%%%%%%%%%%%%%%%%%%%%%%%%%%%%%%%%%%
%
%               BAT SURVEY
%
%%%%%%%%%%%%%%%%%%%%%%%%%%%%%%%%%%%%%%%%%%%%%%%%%%%%%%%%%%%%%%%%%%%%%%%%%%%
\section{Clusters in the {\it Swift}/BAT Survey}
\label{subsec:batsurvey}
The Burst Alert Telescope  \citep[BAT;][]{barthelmy05}, on board the 
{\it Swift} satellite \citep{gehrels04},
 represents
a major improvement in sensitivity for imaging of the hard X-ray sky.
BAT is a coded mask telescope with a wide field of view 
(FOV, $120^{\circ}  \times 90^{\circ}$ partially coded)
sensitive in the 15--200\,keV domain.
As shown in several works \cite[e.g.][]{ajello08a,ajello09b,tueller10},
thanks to the deep exposure, BAT reaches sub-mCrab
(e.g. $<10^{-11}$\,erg cm$^{-2}$ s$^{-1}$) sensitivities in the entire 
high-latitude sky.
This already allowed BAT to detect 10 galaxy clusters above 15\,keV
after $\sim$2 years of all-sky exposure \cite[see][for details]{ajello09a}.
With  $\sim$6\,years of all-sky exposure acquired, the sensitivity of 
BAT has increased substantially leading to the detection
of almost 1000 source in the hard X-ray sky \citep[see ][for details]{cusumano09}.
Here we present a detailed spectral analysis of 10 galaxy clusters
detected in the BAT survey of \cite{cusumano09}.

\begin{deluxetable}{lccccccc}
\tablewidth{0pt}
\tabletypesize{\footnotesize}
\tablecaption{XMM-Newton observations of BAT clusters of galaxies 
\label{tab:xmm}}
\tablehead{
%%%%%%%% column names
\colhead{NAME}  & \colhead{R.A.\tablenotemark{a}}    &\colhead{Decl.\tablenotemark{a}}          & 
\colhead{Date}  & \colhead{OBSID} & \colhead{Exposure\tablenotemark{b}} &
\colhead{Radius\tablenotemark{c}}\\
%%%%
\colhead{}      & \colhead{(J2000)} &  \colhead{(J2000)} & 
\colhead{}      & \colhead{}      & \colhead{(ks)} &\colhead{(kpc)}
}
\startdata
  Abell   85  &   10.4303   &  -9.3483 & 2002-01-07 & 0065140101 & 10.0 & 635.4\\
	 Abell  401  &   44.7395   &  13.5837 & 2002-02-04 & 0112260301 & 13.6 & 823.8\\
  Bullet      &   104.6176  & -55.8974 & 2000-10-21 & 0112980201 & 46.7 & 2460.0\\
  PKS 0745-19  &   116.8758  & -19.3462 & 2000-10-31 & 0105870101 & 28.3 & 1100.0\\
  Abell 1795  &   207.1856  &  26.5928 & 2000-06-26 & 0097820101 & 66.5 & 702.6\\
  Abell 1914  &   216.5013  &  37.8071 & 2002-12-18 & 0112230201 & 25.8 & 1663.8\\
  Abell 2256  &   256.0720  &  78.6301 & 2006-08-04 & 0401610101 & 50.4 & 662.4\\
  Abell 3627  &   243.6066  & -60.8348 & 2004-09-19 & 0204250101 & 22.6 & 204.1\\
  Abell 3667  &   302.9667  & -56.8407 & 2004-05-03 & 0206850101 & 67.3 & 636.1\\
  Abell 2390  &   328.4471  &  17.7516 & 2001-06-19 & 0111270101 & 23.1 & 2079.6\\

\enddata

\tablenotetext{a}{Swift/BAT coordinates are from the work of 
\cite{cusumano09}.}

\tablenotetext{b}{Nominal XMM-Newton exposure before data screening.}

\tablenotetext{c}{Extraction radius in physical units corresponding to
the region of radius 10\arcmin\  around the BAT position used
to extract the XMM-Newton spectrum of the cluster.}

\end{deluxetable}

%%%%%%%%%%%%%%%%%%%%%%%%%%%%%%%%%%%%%%%%%%%%%%%%%%%%%%%%%%%%%%%%%%%%%
\subsection{Analysis of {\it Swift}/BAT Data}
The BAT makes images of the sky thanks to a coded mask (with a random
pattern) placed above a position sensitive detector plane 
\cite[see][ for details]{barthelmy05}. The sky radiation passing through the 
aperture is coded by the mask pattern and recorded in the 
detector plane. The pattern of the mask is such that a source
at a given position in the FOV casts a unique shadow onto the detector
plane and thus its emission can be easily deconvolved. The randomness
of the mask pattern ensures that the cross-talk between sources
(e.g. some flux from a given source is wrongly attributed to another one)
at different positions in the FOV is minimum. Moreover, to minimize
this and other systematic uncertainties that can arise in the BAT survey,
{\it Swift} adopts a random roll-angle strategy when pointing at the
 same position
in the sky. This means that whenever {\it Swift} is pointing at a 
given direction in the sky,  the roll angle (e.g. the angle
on the plane orthogonal to the pointing direction) is chosen
randomly within a range of the nominal (e.g. within $\pm$2\,degrees) 
pointing. This ensures that pointings are never exactly the same
and that sources never fall in the same relative positions
in the BAT FOV.

A decoding procedure is required in order to reconstruct the original sky 
image. A variety of methods can be used to reconstruct the sky image 
in the case of a coded mask aperture
 \citep[see][for a general discussion on reconstruction methods]{skinner87,ajello08a}.
Among them, standard cross-correlation of the shadowgram (e.g. the information
recorded on the detector plane) with a 
deconvolution array, the mask pattern, via fast Fourier transforms (FFTs), 
is the most often used. Generally, sky images are obtained for each 
individual observation, where an observation is defined as a period 
during which the attitude is stable and constant. Subsequently, 
another procedure, such as resampling and re-projecting, is needed 
in order to assemble the final all-sky image.
As  discussed in the Appendix in \cite{ajello08b},
the {\it balanced correlation} \citep{fenimore78}
used to deconvolve BAT observations and source spectra performs
a standard  background subtraction (see the above references
for the exact implementation of this method). However,
this technique works well as long as the background in the array
is flat and not correlated with the mask pattern.
The background in the BAT array
is not flat due to the presence of many background components,
the brightest of which is (below $\sim$60\,keV) the Cosmic X-ray Background
\citep[see][for details]{ajello08c}.
Thus, the balanced correlation alone provides imperfect results
and produces a noticeable background contamination in the 
sky observations and in the source spectra.
This background contamination has been estimated \cite[see][]{ajello08b}
to be $\leq$2\,\% of the Crab Nebula intensity 
in the  14--195\,keV band (e.g. the BAT band). 
Thus, this contamination  does not pose problems for strong sources, but
becomes very relevant for the  (spectral) analysis of faint objects
with $\sim$mCrab intensities as the clusters of this work.
In order to correct for this residual background contamination we 
use the recipe  presented in \cite{ajello08b}.
We use several templates\footnote{For reference see the description
of the {\it batclean} tool available at 
http://heasarc.nasa.gov/lheasoft/ftools/headas/batclean.html.}
 of the BAT background
(for each channel) which are fit together with the contribution
of point sources to the BAT detector counts. In each observation
the residuals are analyzed to check for additional structures
and deviation from Gaussian statistics. If those are found, then
thousands of residual maps are averaged
together (in image coordinates) 
to create a {\it blank field} observation\footnote{Working in
detector coordinates
 ensures that the contribution of any unsubtracted point source is
averaged out.}. These {\it blank field} observations become part
of our template library of background models and fit once again
to any observations which are being used. The process of residuals-inspection 
and template-creation is repeated until the residuals do not show any
systematic feature. For a given observation, the last template model to be added
 is generated from
observations which are close to it in time. This ensures that long-term variation in the BAT background
(due e.g. to the orbit, activation of the spacecraft, noisy pixels etc.)
are correctly taken into account.

Adopting this technique and filtering the data in the way described in
details in \cite{ajello08a}, we extracted a  15--195\,keV  spectrum for 
the 10 galaxy clusters of this analysis. We used all the available
observations at the time of this work 
(approximatively from 2005 to March 2010) resulting in an
 average exposure
at each of the 10 positions  larger than 14\,Ms.
It has already been shown that spectra extracted with this method
are reliable and accurate over the entire energy range 
\citep[see $\S$~2.3 of][for details]{ajello09a}. 

%%%%%%%%%%%%%%%%%%%%%%%%%%%%%%%%%%%%%%%%%%%%%%%%%%%%%%%%%%%%%%%%%%%%%%%%%%%
%
%
%     EXTENDED SOURCES with BAT
%
%
%%%%%%%%%%%%%%%%%%%%%%%%%%%%%%%%%%%%%%%%%%%%%%%%%%%%%%%%%%%%%%%%%%%%%%%%%%%
\subsection{Studying ``Extended'' Sources with {\it Swift}/BAT}
\label{sec:ext}
Coded mask telescopes are designed and optimized for the study
of point-like sources \citep[e.g. see ][ and references therein]{ajello08c}. 
Formally the mask acts as a filter,
canceling out those signals (celestial and 
not) whose spatial frequency is larger than the spatial frequencies 
of the mask tiles. This means that if an X-ray source extends over an area
which is larger than the projection of the mask tile on the sky
(i.e., 22.4\arcmin\ or, which is the same,
 the Full-Width at Half Maximum, FWHM, of the BAT
Point Spread Function, PSF) then part of the X-ray flux is {\it necessarily
lost} in the background.  In simpler words, if a source is extended, then
its shadow (produced when the radiation passes through the mask)
on the detector plane looses part of its contrast  (i.e., the mask
is illuminated from all sides). The limiting case is represented
by the Cosmic X-ray Background, which extending over the entire
sky, is completely removed by the BAT mask. Indeed in order
to measure the Cosmic Background,  different, non-standard, techniques have
to be used \citep{ajello08c}.

Clusters of galaxies are X-ray sources extending up to $\sim$1\,degree.
In our first work \cite{ajello09a}, we showed that all the clusters
detected by BAT are consistent with being point-like sources (for BAT)
with the exception of Coma which is clearly resolved. Given the properties
of coded masks expressed above, it is a good thing that clusters
are seen as point-like sources. Indeed, in this case the flux measurement
is correct while it is not if the source is {\it detected} as extended.

\begin{figure}[h!]
\begin{centering}
	\includegraphics[scale=0.7]{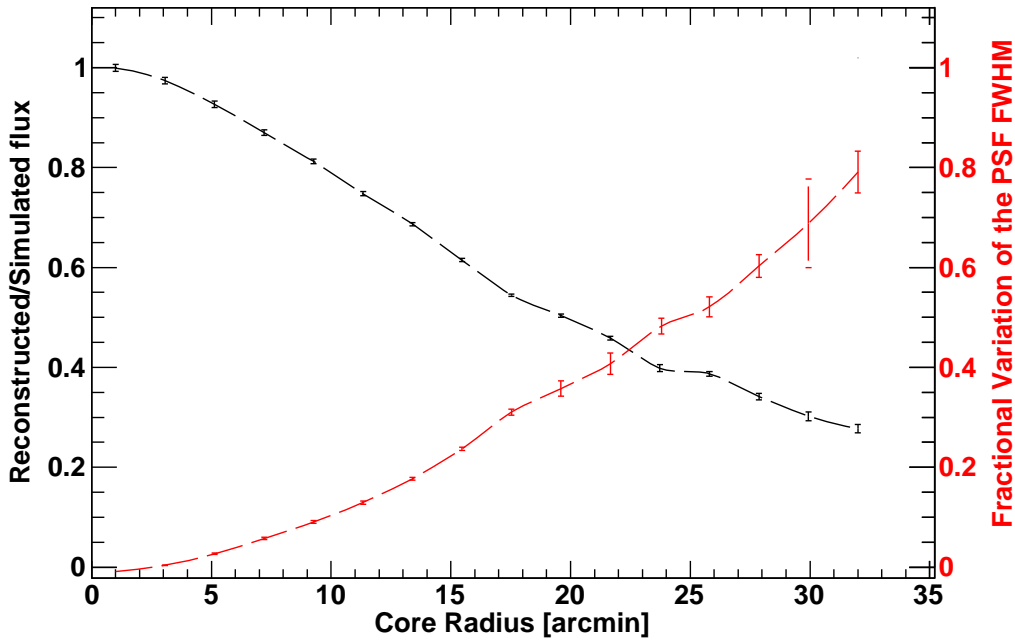} 
	\caption{Suppression of the source flux (solid line)
due to the source extension
as a function of core radius ($r_c$) for an emission profile that follows
a beta model (see $\S$~\ref{sec:ext}
for details). Error bars are of statistical origin and reflects
the uncertainty in the reconstructed quantities.
The dashed line shows the fractional increase of the FWHM
of the source PSF (with respect to the PSF for a point-like source) when the
source is ``resolved'' by BAT. For a cluster with a core radius of
 30\arcmin\ the FWHM is $\sim$30\,\%
larger than the point-like source FWHM. 
	\label{fig:ext}}
\end{centering}
\end{figure}

In order to quantify this effect in more detail
 we performed several  Monte Carlo simulations.
We simulated an extended source whose surface brightness profile can
be approximated by a {\it beta model} of the form:
\begin{equation}
F(r) \propto \left [ 1 + \left(\frac{r}{r_c}\right )^2 \right]^{0.5 -3\beta}
\end{equation}
where $r_c$ is the core radius and $\beta$ is typically in the 0.5--0.8 range
\citep[see e.g.][]{ettori98,ajello09a}.
We then reconstructed the sky image and  detected the simulated source
measuring its flux and  the PSF FWHM. Simulations were  performed, in the 15--55\,keV band\footnote{The results of this analysis do not change if a different band is chosen.},
 for several core radii and for a fixed $\beta$ of 0.6.
The results of this exercise are reported in Fig.~\ref{fig:ext}.
It is clear that a substantial part of the source flux is lost
if clusters have a core radius larger than 5\arcmin\ .
This effect is a function of the emissivity profile of the source
and depends on it. If we were considering a source with a Gaussian
emission profile or a  uniform profile (e.g. a disk-like emission) than
the flux suppression would be even stronger.
The presence of a cool-core acts in the opposite direction, indeed
in this case a large part of the cluster's emission is confined
within the inner few arc-minutes from the core 
(normally within the core radius) where the flux suppression (due
to source extension) is negligible.
Moreover, Fig.~\ref{fig:ext} gives a powerful tool to understand
when the flux suppression takes place. Indeed, in all these
cases the FWHM of the source is larger than the point-like one (i.e.
22.4\arcmin\ ). Coma which is resolved by BAT can be used as an example.
Indeed, 
adopting a core radius of $r_c=10.7$\arcmin\ \citep[see e.g.][]{lutovinov08},
we can estimate from Fig.~\ref{fig:ext} that the measured FWHM should be
$\sim26$\arcmin\ . This is found in good agreement with the results
reported in \cite{ajello09a}. Moreover from the same graph we can estimate
that the flux suppression is $\sim$25\,\%.
The next most extended  clusters reported in \cite{ajello09a}
are Ophiucus and Perseus.
Ophiucus with a core radius of 3.2\arcmin\ \citep{wat01} is not
resolved by BAT and thus the flux suppression is negligible.
Perseus has a core radius of 4.7\arcmin\ \citep{ettori98} and thus
at the limit where the flux suppression might start playing a role.
However, Perseus has a bright cool core whose emission profile can
be modeled with a power law \citep[e.g. see ][ and references therein]{ettori98}. This concentrate most of the cluster's emission in the core which is never
resolved by BAT. Hence, Perseus is detected as a point-like object.
The majority of the clusters detected by BAT, either those reported here or
those described in \cite{ajello09a}, are detected as point-like objects.
Norma (reported in this work) with a core radius of $\sim$10\arcmin\
like   Coma, is also resolved by BAT. However, its 
marked North-West elongation and the presence of a bright nearby AGN
make its case more complex that the 
simple spherically symmetric case discussed here. This,
will discussed in details in $\S$~\ref{sec:norma}.

%Finally we note that there exist methods to correctly treat and analyze
%the emission from resolved sources
%with coded mask \cite[e.g.][]{renaud06}. However their applications
%to BAT requires a level of modeling of the instrumental response
%which is outside of the scope of this paper. 

To summarize this section, we note that coded-mask telescopes are optimized
for the study of point-like objects and can be used for the study
of (intrinsically) extended objects only if these are detected as point-like
(e.g. for a beta model this is true if the core radius is less than 5\arcmin\ ).
 If an object is detected as extended,
then part of its flux (depending on the source brightness profile) is 
suppressed. The deviation of the PSF from the point-like PSF can be
used to understand whether this effect is present. Among all the clusters
detected by BAT, Coma and Norma (see next sections)
are the only two sources which are {\it extended} in 
BAT and part of the source flux is lost in the
BAT background. All the other clusters detected by BAT
are not resolved by BAT and BAT can
be safely used for the study of their emission.

%%%%%%%%%%%%%%%%%%%%%%%%%%%%%%%%%%%%%%%%%%%%%%%%%%%%%%%%%%%%%%%%%%%%%%%%
%
%             XMM 
%
%%%%%%%%%%%%%%%%%%%%%%%%%%%%%%%%%%%%%%%%%%%%%%%%%%%%%%%%%%%%%%%%%%%%%%%%
\subsection{Analysis of XMM-Newton Data}
\label{sec:xmm}
For all the clusters, we extracted a 0.5--8.0\,keV spectrum using
publicly available XMM-Newton (EPIC-PN) observations. 
The details of these observations are reported in Tab.~\ref{tab:xmm}. 

For each observation, 
XMM-Newton data are screened filtering for periods of flaring background.
This is done examining the lightcurve in the 10-12\,keV band and
determining the rate of the quiescent background (the rate
outside of the flaring episodes).
For all the observations reported in this analysis this was very close to
0.5\,counts s$^{-1}$ in agreement with 
e.g. \cite{nevalainen05} (and references therein).
However, this is not sufficient for filtering out flaring episodes that
produce a soft background component \citep[e.g.][]{nevalainen05,carter07}.
We thus inspect the lightcurve in the 1--5\,keV band extracted in an annulus
of inner radius 10\arcmin\ and outer radius  12\arcmin\ .
We filter out all those times bins that deviate more than 3\,$\sigma$
from the average quiescent background
(determined through a Gaussian fit to the histograms of the rates).
 
We extracted
the cluster spectrum using a single extraction region with a radius of
$\sim$10\arcmin\ . This is partly motivated by: 
i) the extent of the XMM-Newton EPIC-PN CCD ($\sim$12\arcmin\ in radius),
 ii) the fact that we use the two outer arcminutes to perform background filtering and 
iii) the fact that 
for a typical beta profile (e.g. core radius of 3.8\arcmin\ and $\beta$=0.7)
this selection includes up to 94\,\% of the entire cluster emission and more than
that if the cluster has a cool core. The fact that the BAT PSF is consistent
with the point-like one for all the clusters in this analysis (except Norma)
suggests
that the hard X-ray emission (above 15\,keV) is coming from the inner part
of the cluster. If this were not the case, we would have
observed a significant deviation in the FWHM of the BAT PSF.
%In this framework it is justified to extract the cluster emission within
%the inner $\sim$10\arcmin\ .
Tab.~\ref{tab:xmm} shows the dimension in physical units (e.g. kpc) 
of the extraction radius of 10\arcmin\ at the redshift of the source.

The level of the background was evaluated using blank-sky observations\footnote{Blank-observations are described and made available at: http://xmm.vilspa.esa.es/external/xmm\_sw\_cal/background/blank\_sky.shtml\#BGfiles.} 
\citep[e.g. see][]{lumb02,read03,nevalainen05} which are described
in details in \cite{carter07}. The black fields were selected
from the same sky region as the observation under analysis and 
with a similar foreground absorbing column density.
These fields were then reprojected to the source (sky) coordinates
system and processed in a similar way as the observation under analysis.
For each cluster, a background spectrum has been extracted from
the exact same 10\arcmin\ region as the cluster.
To allow for different intensities of the background components
(between the source and the 'background' observations) we renormalized
the background spectrum by the ratio of the total emission in the
10--12\,keV band in the annulus with inner and outer radii of 10\arcmin\
and 12\arcmin\ respectively \citep[see also][]{molendi09}.

The results of the background subtraction change slightly by
varying this renormalization constant within its error, 
as well as changing the extraction
annulus of the background or the thresholds for the removal of the 
flaring episodes (in the hard and the soft bands). 
We noticed that the spectral
results are robust for variation of the aforementioned parameters
%(and the observations reported in Tab.~\ref{tab:xmm}) 
if a systematic
uncertainty of $\sim$2\,\% is applied to our background subtracted spectra.
We thus employed this systematic
uncertainty connected to the background subtraction, in the 0.5--8\,keV band,
when fitting the XMM-Newton data.
Finally, all the spectra  were rebinned in order to have
a minimum of 50 counts ($\geq$7\,$\sigma$) per bin.

\begin{figure}[h!]
\begin{centering}
	\includegraphics[scale=0.6,angle=270]{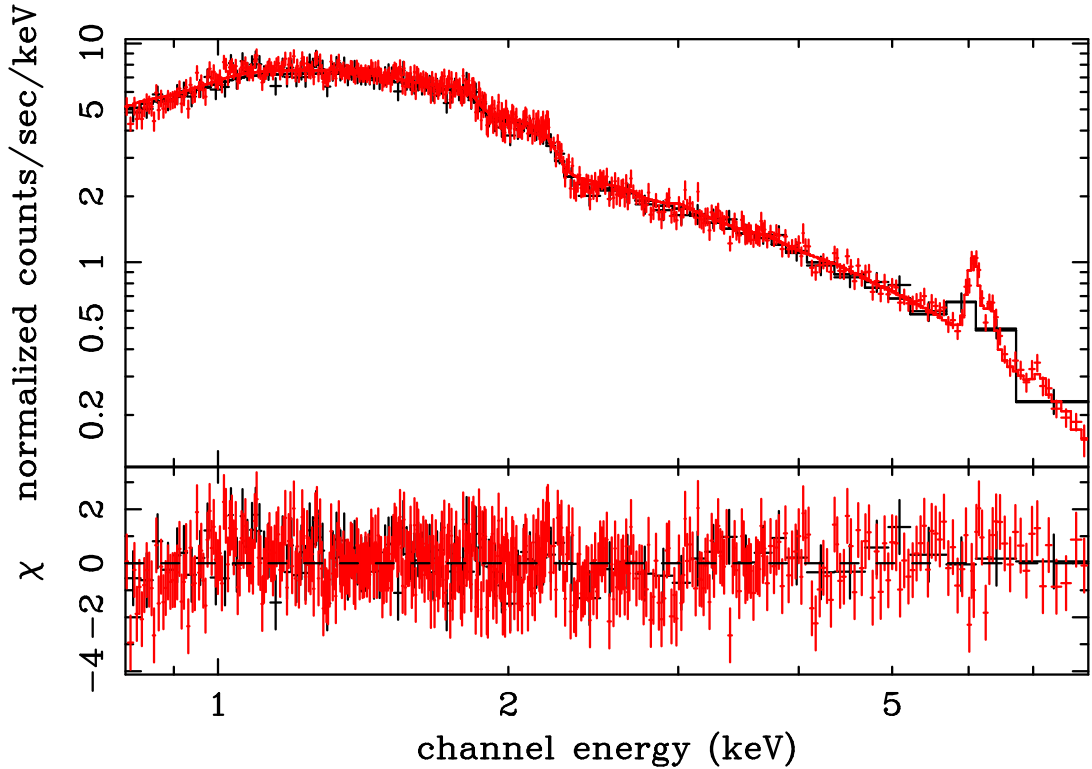} 
	\caption{Spectra of PKS 0745-19 using two different background
subtraction techniques: blank field observations (in red) and
an extraction region free of cluster's emission (in black).
The spectra were rebinned differently to simplify the comparison.
	\label{fig:bkg}}
\end{centering}
\end{figure}

The most distant clusters in our sample provide a  test bed for 
checking the goodness of the background subtraction employed in this work.
Indeed, for the Bullet cluster, PKS 0745-19 and Abell 2390 is possible to
find spatial 
regions of the EPIC-PN CCD that are the least contaminated by the cluster
emission. These regions provide a clean way to determine the background spectrum
which suffers from different systematic uncertainties\footnote{The main
systematic uncertainty is that the background spectrum is extracted
from a region different from the region used to extract the cluster's spectrum.}
 with respect to the use of blank fields. 
In all these cases we found that within
the aforementioned systematic uncertainty, the two background subtracted spectra
(e.g. the one that uses blank field observations and the one that uses
part of the CCD which is not contaminated by the cluster emission) are 
in good agreement. One such example is reported in Fig.~\ref{fig:bkg}, which
shows the two background subtracted spectra (generated with the two
different techniques described above) of PKS 0745-19 fitted with a single
temperature thermal model. As it is apparent from this figure, there is 
very good agreement between the two spectra.

%%%%%%%%%%%%%%%%%%%%%%%%%%%%%%%%%%%%%%%%%%%%%%%%%%%%%%%%%%%%%%%%%%%%%%%%
%
%             XMM  Point-like sources
%
%%%%%%%%%%%%%%%%%%%%%%%%%%%%%%%%%%%%%%%%%%%%%%%%%%%%%%%%%%%%%%%%%%%%%%%%

\subsection{Point-like Sources in the XMM-Newton Fields}
\label{sec:sources}
Given the extent of its PSF, BAT is unable to discriminate
the cluster's emission from that of nearby point sources which fall in
the cluster's region.
Thus, if present these sources would
 contaminate with their signals the cluster's
emission as seen by BAT. For this reason, when analyzing
XMM-Newton data jointly with BAT, we do not filter out the point sources which
are clearly resolved and  detected  by XMM-Newton.
However, we employ two different approaches to determine
whether these sources contaminate in any way the total X-ray signal
in the BAT band.

In the more general approach we evaluate the contamination
from point sources using the 2XMM catalog \citep{watson09}
which contains all point sources detected in the XMM-Newton
fields including the observations used in this work (e.g. \ref{tab:xmm}).
From the 2XMM catalog we select
all the point-sources detected in the 0.2-12 keV band,
 with a likelihood 
probability of being spurious -ln(P)$>$15 and within 10\arcmin\
from the {\em Swift}-BAT centroid. The count rate of all the sources
in the field of view has been summed and converted by extrapolation 
into 15--55\,keV fluxes in units of erg cm$^{-2}$ s${-1}$.
In this conversion we assumed 
that the spectra of these sources were represented by a power-law with a
photon  index of $\Gamma$=2, absorbed by a
cold material with column-density N$_H$ equal to the Galactic value.
We believe that this approach will yield conservative estimates for the
contamination to the 15--55\,keV signal produced by point-like sources.
Indeed, by assuming a power-law spectrum we implicitly assumed that
all sources detected by XMM-Newton are AGN, while it is known that
 a large fraction
of the sources detected in clusters' region show a thermal spectrum
\citep[see Fig.~10 in][ and references therein]{finoguenov04} 
and thus will have a negligible $>$10\,keV emission.
As shown in \cite{burlon10}, the average broad-band 1--200\,keV
{\it intrinsic}\footnote{Photoelectric absorption caused by  the
circumnuclear material around the source will make the AGN spectra look
harder in the $<10$\,keV band, but the $>10$\,keV continuum will be mostly
unaffected  \citep[see][for details]{burlon10}.} 
spectra of AGN is compatible with a power-law with a photon-index of $\sim$2.0.
Our  estimates are reported in Tab.~\ref{tab:sources}. 
For most of the clusters in this analysis, the estimated
contamination from point-sources is a factor $\sim50$ below the total
(cluster plus sources) emission measured by BAT (see Tab.~\ref{tab:spec} for
details),  and thus negligible.
For only three clusters, PKS 0745-19, Abell 1795, and Abell 2390
 the contamination might
be relevant (still a factor $\sim5$ below the total flux).

In the second approach, we study the XMM-Newton observations closely and
we extract the spectrum of the brightest sources (up to 5)  in
the field. The characterization of their spectra in the 0.5--8.0\,keV
band allows us to make a solid prediction (i.e. without assumptions) 
of the contaminating signal in the BAT band. 
The findings will be discussed case by case in the
next sections.
%but in no case the emission from point sources was found
%to be comparable with the one of the cluster.

\begin{deluxetable}{rcccc}
\tablewidth{0pt}
\tabletypesize{\footnotesize}
\tablecaption{Summed X-ray emission from all the point-sources detected
by XMM-Newton in the fields of the  clusters.
\label{tab:sources}}
\tablehead{

\colhead{NAME}  &  \colhead{Count-rate\tablenotemark{a}} & 
\colhead{\# Sources}     &  \colhead{Flux$_{\rm 0.2-12\,keV}$}
 &  \colhead{Flux$_{\rm 15-55\,keV}$}\\
\colhead{}      & \colhead{count s$^{-1}$}   &
\colhead{}      & \colhead{ erg cm$^{-2}$ s$^{-1}$}
& \colhead{ erg cm$^{-2}$ s$^{-1}$}\\
}
\startdata

Abell 85     & 0.23 & 10 &  5.61$\times$10$^{-13}$ &  1.78$\times$10$^{-13}$\\
Abell 401    & 0.10 & 4  &  3.72$\times$10$^{-13}$ &  1.18$\times$10$^{-13}$\\
Bullet       & 0.10 & 4  &  2.38$\times$10$^{-14}$ &  7.55$\times$10$^{-14}$\\
PKS 0745-19  & 0.63 & 14 &  4.88$\times$10$^{-12}$ &  1.55$\times$10$^{-12}$\\
Abell 1795   & 1.76 & 13 &  4.28$\times$10$^{-12}$ &  1.36$\times$10$^{-12}$\\
Abell 1914   & 0.34 & 19 &  8.07$\times$10$^{-13}$ &  2.56$\times$10$^{-13}$\\
Abell 2256   & 0.21 & 10 &  4.98$\times$10$^{-13}$ &  1.58$\times$10$^{-13}$\\
Abell 3627   & 0.15 & 1  &  3.56$\times$10$^{-13}$ &  1.13$\times$10$^{-13}$\\
Abell 3667   & 0.37 & 27 &  6.90$\times$10$^{-13}$ &  2.19$\times$10$^{-13}$\\
Abell 2390   & 0.51 & 16 &  12.1$\times$10$^{-13}$ &  3.85$\times$10$^{-13}$\\

\enddata
\tablenotetext{a}{Summed count rate of all sources detected by XMM-Newton
in the 0.2--12\,keV band.}

\end{deluxetable}

%%%%%%%%%%%%%%%%%%%%%%%%%%%%%%%%%%%%%%%%%%%%%%%%%%%%%%%%%%%%%%%%%%%%%%%%
%
%             XMM  + BAT
%
%%%%%%%%%%%%%%%%%%%%%%%%%%%%%%%%%%%%%%%%%%%%%%%%%%%%%%%%%%%%%%%%%%%%%%%%
\subsection{Joint Analysis of XMM-Newton and {\it Swift}/BAT Data}
\label{sec:xmmbat}

Spectral analysis of  XMM-Newton and {\it Swift}/BAT data
has been performed, for all the clusters, using XSPEC 
\citep[version 12.5.1 in][]{arnaud96}.
Observations of the Crab Nebula showed that, in principle, the inter-calibration
of the two instruments is good within $\sim5$\,\%.
Indeed, as reported in \cite{kirsch05}, the Crab Nebula 0.3--10\,keV
spectrum as observed with the EPIC-PN can be modeled
as an absorbed power-law with a photon index of 2.125, a normalization
of 8.86 and an absorbing column density of 4.08$\times 10^{21}$\,atoms cm$^{-2}$.
For BAT\footnote{See http://swift.gsfc.nasa.gov/docs/swift/analysis/bat\_digest.html for details.}, the Crab Nebula 15-200\,keV spectrum is compatible with
a power-law with a photon index of 2.15 and normalization of 10.17.
Thus, for the 15-55\,keV band (where most of the clusters' signal
is concentrated for BAT) the two Crab Nebula parametrizations
yield a flux of 1.21$\times10^{-8}$\,erg cm$^{-2}$ s$^{-1}$ and 
1.28$\times10^{-8}$\,erg cm$^{-2}$ s$^{-1}$ for XMM-Newton and BAT respectively.
Thus, we expect the inter-calibration of the two instruments to be close
to unity (within $\sim$5\,\%). However, one must take into account
that the Crab Nebula is a very bright target for XMM-Newton and in order
to avoid pile-up problems the observations presented in \cite{kirsch05}
were performed in ``Burst Mode'' \cite[see ][for details]{kirsch05} rather
than the Full-Frame Mode normally used for studying diffuse sources
with XMM-Newton. Thus, the same inter-calibration between XMM-Newton and BAT
might not necessarily apply in this case. However, \cite{burlon10} performed
spectral fitting of 12 faint  AGN using XMM-Newton (with the EPIC-PN in
Full-Frame Mode) and {\it Swift}/BAT. In all these cases the inter-calibration
between the two instruments has been found to be compatible with unity.
As a strategy in the spectral fitting presented in the next section, we
employed a normalization constant to take
into account differences in the calibrations of the XMM-Newton and BAT
instruments. This constant has been fixed to 0.95 to take into
account the different Crab Nebula spectra as observed with the two
instruments. However we performed some tests with the joint datasets
of this paper and found out that changing the inter-calibration constant
by $\pm10$\,\% produces a negligible change (e.g. less than 1\,\%)
in the best-fit temperatures and their uncertainties. 
Thus the
results that will be presented are robust against variation of the
aforementioned inter-calibration constant.
The reason for the small variation of the temperature with the
 inter-calibration constant lies in the different signal-to-noise
ratios of the two datasets. Indeed, in a joint fit the best-fit temperature
is entirely constrained by the signal in the XMM-Newton band 
and a small variation of the inter-calibration constant (e.g.
moving the less significant 
BAT data up or down around the best fit) does not change the results.
We also checked that leaving this constant free to vary did not
produce any appreciable improvement in the fit (in terms of goodness
of fit) for all the clusters presented in this work.

We started fitting all the spectra with a single-temperature thermal model (APEC)
with absorption fixed at  the Galactic value.
Only if the value of the $\chi^2/dof$ was significantly
greater than 1, we tried
to add a second thermal model or a power law. In this case
we chose the model which produced the best improvement in the fit
(evaluated using the F-test) and the best residuals.
In all spectral fits  all the parameters are tied together within
the two datasets (e.g. XMM-Newton and BAT).

In order to test which is the maximum level of non-thermal
emission which is allowed by our data, a power law
has been added to the best-fit model of every cluster.
The power-law index has been fixed to 2.0,
which is a value generally accepted for the non-thermal
hard X-ray
component generated by IC of relativistic  electrons off CMB photons
\cite[e.g.][]{reimer04,nev04}. 
We then let the power-law normalization vary until the $\Delta \chi^2$
increment was larger than  2.7(6.64). 
According to \cite{avni76}, this gives the 90\,\% (99\,\%) confidence
level on the parameter of interest. This allows us to investigate
the level of non-thermal flux which is consistent with our data.
In the next sections, the details of the spectral analysis of each
single cluster are reported.

\section{Spectral Analysis of Individual Clusters of Galaxies}
%%%%%%%%%%%%%%%%%%%%%%%%%%%%%%%%%%%%%%%%%%%%%%%%% Abell 0085
\subsection{Abell 0085}
Abell 85, a galaxy cluster at z=0.0521, 
has been detected for the first time in
 the X-ray band by  Ariel V, thus it is one of the first galaxy clusters
ever detected in X-rays \citep{mitchell79,mchardy78}.
A detailed analysis performed with {\it Einstein} \citep{jones1999} 
showed that the ICM temperature is in the   7--9\,keV range. 
ROSAT observations  \citep{prest} revealed 
that the temperature and surface brightness structures
of A85
are not regular, implying that the cluster is dynamically disturbed. At
the same
time the high central gas density indicates the presence of a cool core.
\citet{enss} showed that A85
has recently experienced a major merging, as indicated  by the presence of 
a radio relic.  The estimate of the magnetic field intensity 
is, for the radio relic, of the order of B$\sim$2.6$\,\mu$G \citep{enss}.
\cite{slee01}, using high-resolution radio observation, determined
that the flux and spectral index of the relic 
at 1.425\,GHz are S$_{\rm R}$=40.9\,mJy and $\alpha=$3.

The combined XMM-Newton and {\it Swift}/BAT dataset when
fit by a single thermal model produces a $\chi^{2}/dof=738.8/622$,
leaving unsatisfactory residuals at high energy.
We thus added an additional thermal model to the fit which
produces a significant improvement in the fit ($\Delta \chi^2\approx 140$ for 3
additional parameters).
The best fit ($\chi^2/dof$= 602.1/619)
temperatures are 6.09$^{+0.43}_{-0.29}$\,keV and 
1.72$^{+0.32}_{-0.06}$ \,keV,  while the respective metallicities are:
0.33$^{+0.04}_{-0.03}$ and 0.15$^{+0.04}_{-0.03}$. 
The low temperature component accounts for the cool core of the cluster.
Our results are in good agreement with the one reported by \cite{durret05}
using XMM-Newton data alone.

\begin{figure*}[ht!]
  \begin{center}
  \begin{tabular}{cc}
    \includegraphics[scale=0.4]{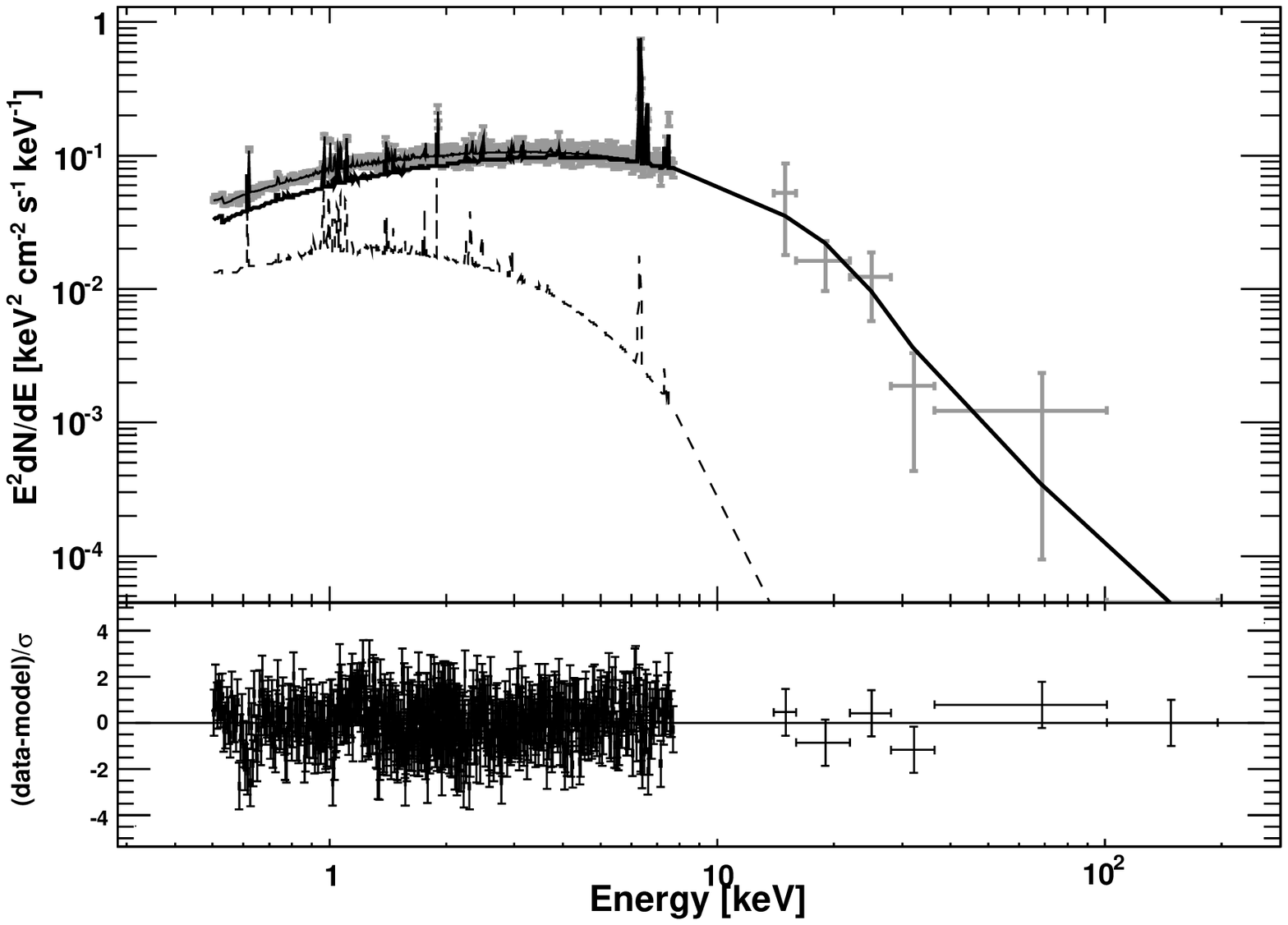} 
 \includegraphics[scale=0.4]{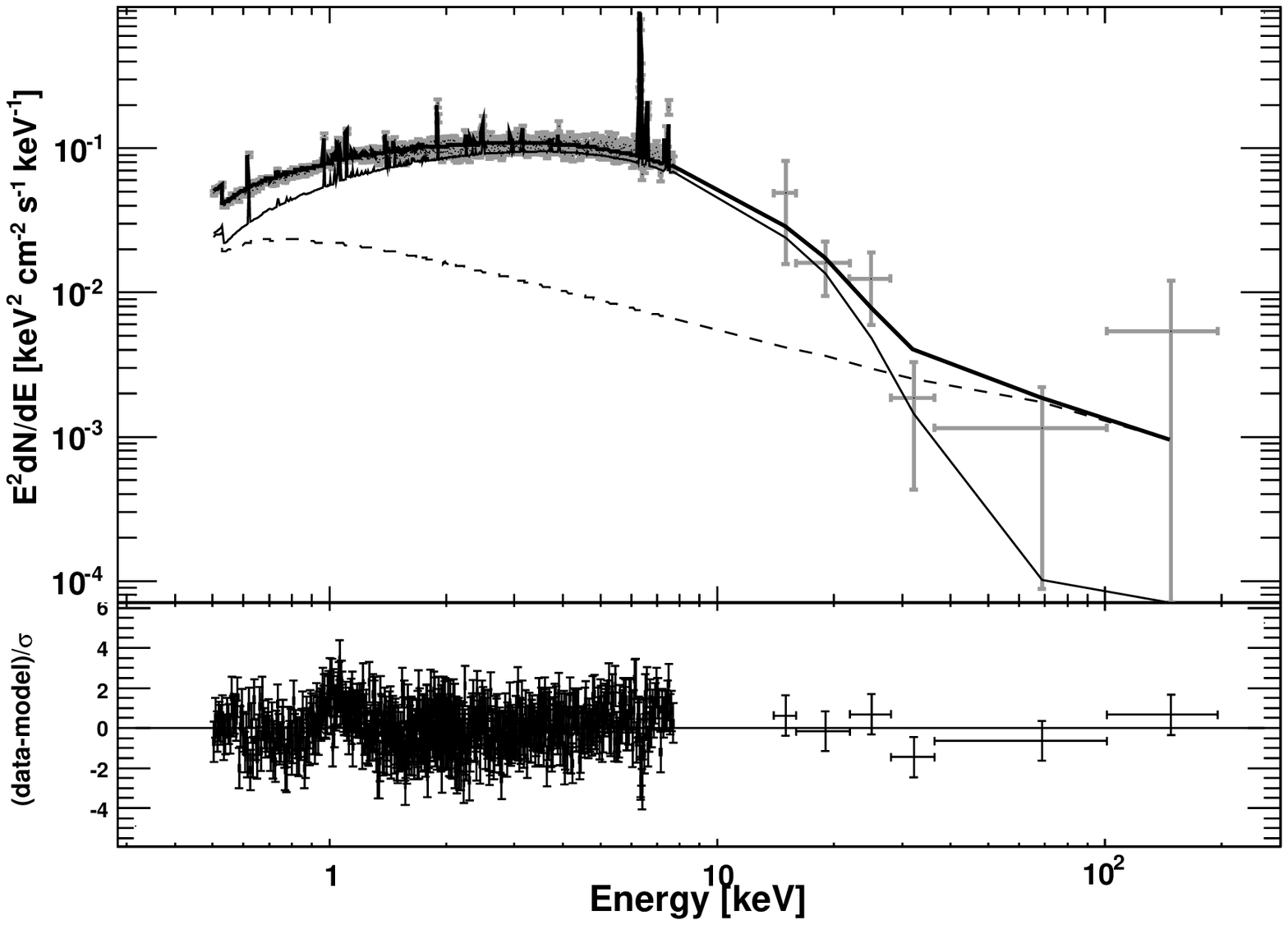} 

\end{tabular}
  \end{center}
  \caption{{\bf Left Panel:}
Spectrum of Abell 85 fitted with the sum of two thermal
models.
{\bf Right Panel:} Spectrum of Abell 85 fitted with the sum
of a thermal model and a power law.
}
  \label{fig:a85}
\end{figure*}

We also tried to add a power law model to the single thermal model.
The fit improves with respect to the single temperature thermal model
with a $\Delta chi^2=59$ for 2 additional degrees of freedom.
The best-fit temperature becomes in this case 5.09$^{+0.14}_{-0.16}$\,keV
and the photon index of the power law 2.68$^{+0.19}_{-0.13}$.
However, the $\Delta \chi^2$ is noticeable larger when
the sum of two thermal model is fit to the data and we consider this model
to be the best representation of our dataset 
(parameters reported in Tab.~\ref{tab:spec}).
Both spectral fits are reported in Fig.~\ref{fig:a85}.

We derive a 99\,\% upper limit on any non-thermal component
of 2.51$\times10^{-12}$\,erg cm$^{-2}$ s$^{-1}$ in the 50--100\,keV band.
The upper limit on the non-thermal luminosity in the 20--80\,keV
band is 3.62$\times10^{43}$\,erg s$^{-1}$. When converted
to the  cosmology used by \cite{nev04} (H$_0$=50\,km s$^{-1}$
Mpc$^{-1}$, q$_0$=0.5 and $\Lambda =0$) this becomes  
6.68$\times10^{43}$\,erg s$^{-1}$ which is in agreement with the
value of 10.7$^{+6.3}_{-6.3} \times10^{43}$\,erg s$^{-1}$ reported
in \cite{nev04}.

%%%%%%%%%%%%%%%%%%%%%%%%%%%%%%%%%%%%%%%%%%%%%%%%% Abell 041
\subsection{Abell 401}
 
Abell 401, at z=0.074, is part of a cluster pair with Abell 399 which
is in a pre-merging state 
\citep[e.g. ][and references therein]{kara80,fujita96}.
It is a rich cluster with a temperature of the ICM
in the 7--8\,keV range \citep{fujita96}. Recently, 
using XMM-Newton,  \cite{sakelliou04}
found an average ICM temperature of 7.23$^{+0.17}_{-0.21}$\,keV.
Abell 401 was one of the first clusters, along with Coma, that
were discovered to host an extended radio emission \citep{har74}.
This radio halo was recently confirmed by deep VLA observations 
\citep{bac03}. The intensity at 1.4\,GHz was found 
to be $S_R=17\pm1$\,mJy.

\begin{figure*}[ht!]
  \begin{center}
  \begin{tabular}{cc}
    \includegraphics[scale=0.4]{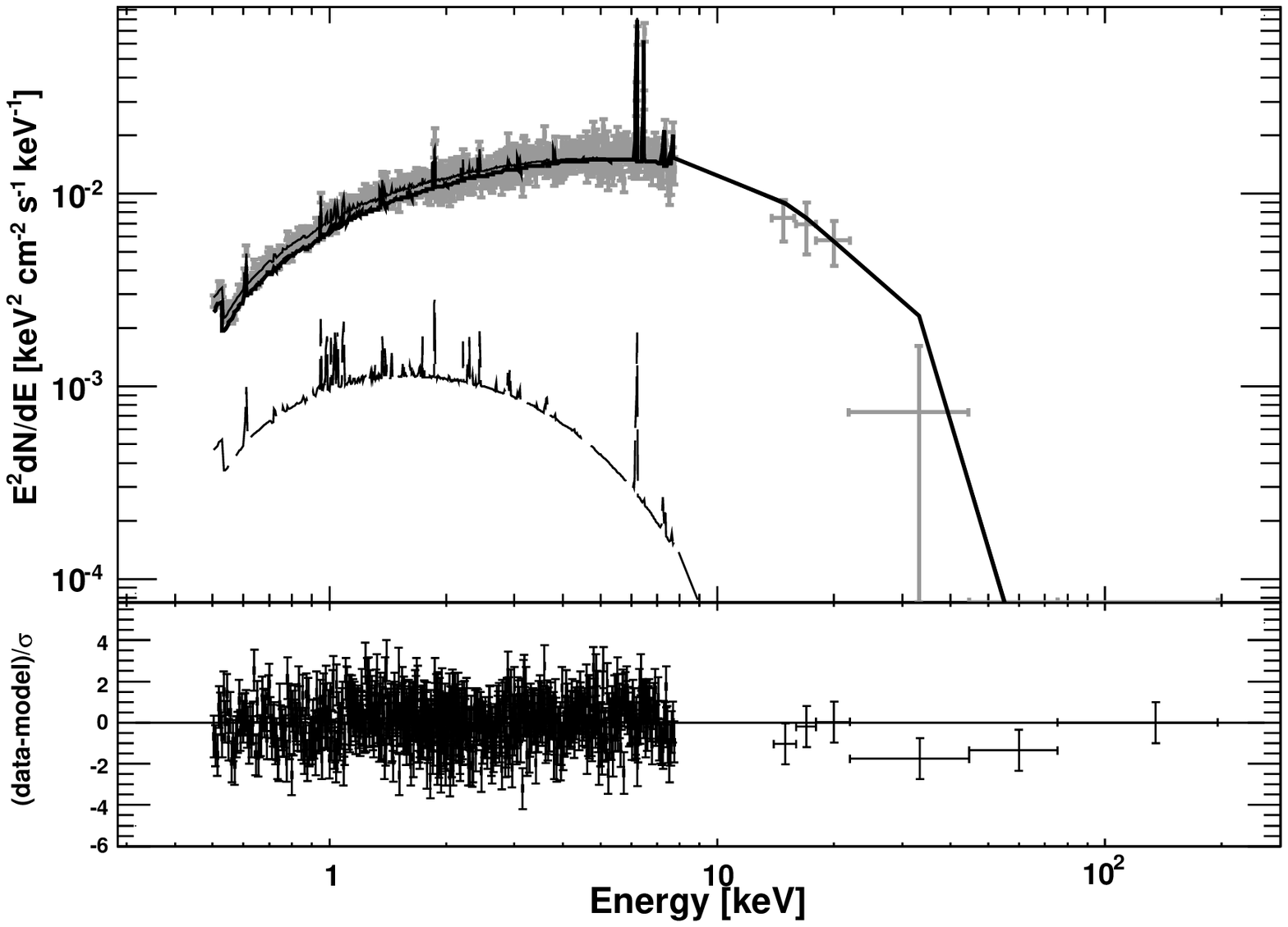}%{aco401_2T.eps} 
 \includegraphics[scale=0.4]{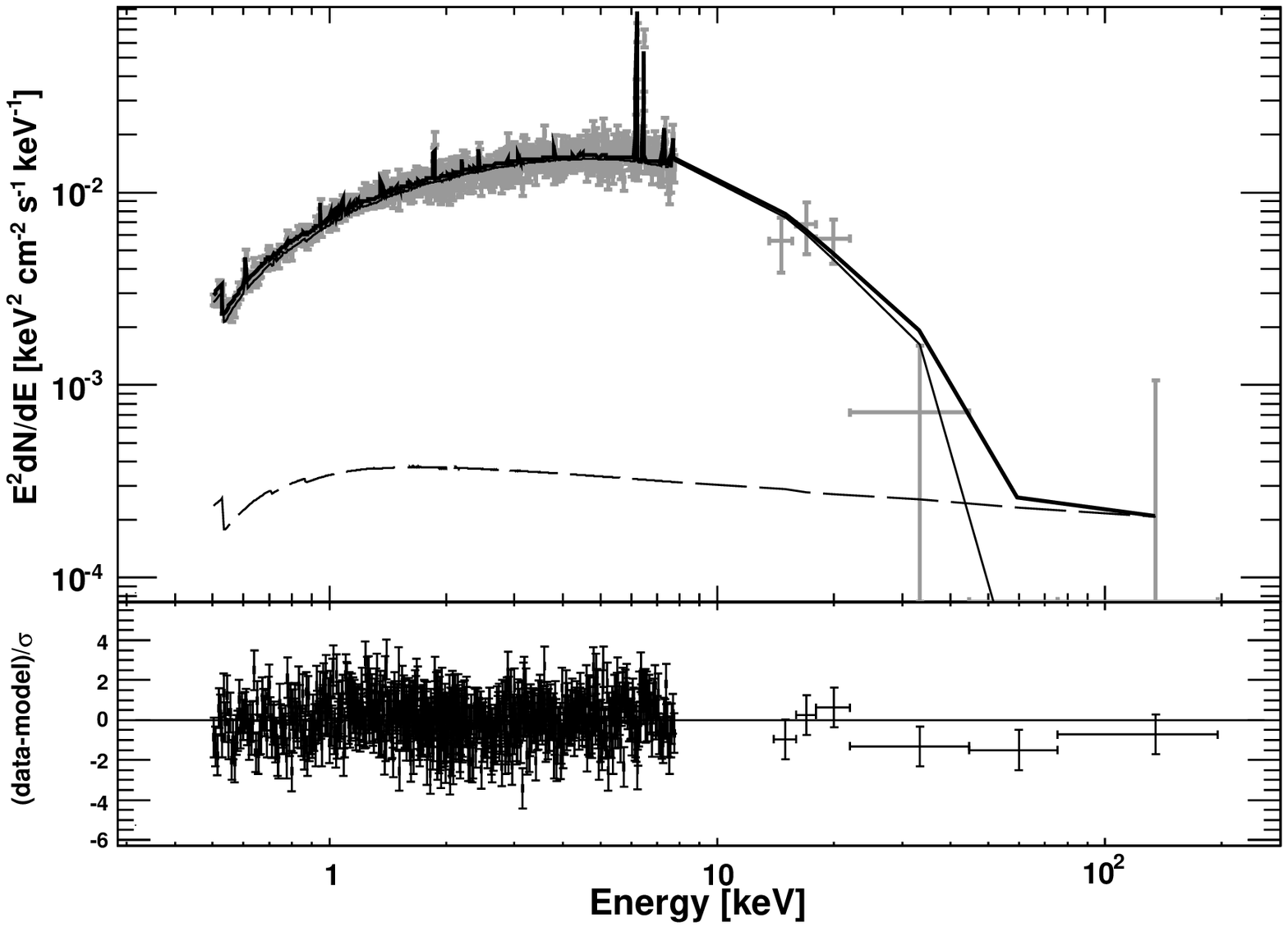}%{aco401_pow.eps} 

\end{tabular}
  \end{center}
  \caption{{\bf Left Panel:}
Spectrum of Abell 401 fitted with the sum of two thermal models.
{\bf Right Panel:} Spectrum of Abell 401 fitted with
the sum of a thermal model and a power law.
}
  \label{fig:a41}
\end{figure*}

The BAT spectrum is well fit by a bremsstrahlung model
with a plasma temperature of 7.79$^{+5.30}_{-2.86}$\,keV.
The combined XMM-Newton-BAT dataset is reasonably well fit
($\chi^2/dof$=766.3/655)
by a single thermal model with a temperature of 7.19$\pm0.17$\,keV
and an abundance of 0.25$\pm0.03$ solar.
Still, adding a second thermal model improves the fit substantially
($\Delta \chi^2=33.9$ for 3 additional parameters).
The temperature and abundance of the hot component are respectively
8.61$^{+0.60}_{-0.46}$\,keV and 0.30$^{+0.04}_{-0.04}$.
Those of the cold  component are  respectively
2.05$^{+0.65}_{-0.45}$\,keV and 0.16$^{+0.11}_{-0.08}$.
This fit is shown in the left panel of Fig.~\ref{fig:a41}.

We also tried a fit with the sum of a thermal model and 
a power-law (see right panel of Fig.~\ref{fig:a41}).
The best-fit temperature and power-law photon index are respectively
7.55$^{+0.26}_{-0.27}$\,keV and 2.16$^{+0.63}_{-0.26}$.
The improvement in the goodness of fit, with respect to the single thermal
model, is $\Delta \chi^2=$9.1 for 
2 additional parameters and is clearly not a better fit than the 
sum of two thermal models. Indeed the F-test yields a probability
of 0.02 and 1.7$\times10^{-6}$  for the power-law and the additional thermal
model respectively of being spurious.
For this reason we believe that the sum of two thermal model is a more
adequate representation of the XMM-Newton/BAT dataset and we report
its  best-fit parameters in Tab.~\ref{tab:spec}.

Since no spectral index for the radio emission is available in the literature
we adopt a value of $\alpha=$2.0.
Using a power-law with a photon index of 2.0 we derive that the 99\,\% CL
upper limit on the non-thermal component in the 50--100\,keV band is 
2.2$\times10^{-13}$\,erg cm$^{-2}$ s$^{-1}$.
As a final note, the brightest point-source in the XMM-Newton 10\arcmin\ 
region is located at R.A.(J2000) = 02:59:05.5 Decl.(J2000) = 13:39:44.9. 
Its spectrum is very
soft and consistent with a bremsstrahlung model with a temperature
of 0.3$\pm0.1$\,keV. Its flux in the 2--10\,keV band is 1.9$\times10^{-16}$\,erg
cm$^{-2}$ s$^{-1}$ and thus it is negligible when compared to the cluster
signal in both the XMM-Newton and BAT bands.

%%%%%%%%%%%%%%%%%%%%%%%%%%%%%%%%%%%%%%%%%%%%%%%%% BULLET cluster
\subsection{Bullet cluster}
  
1E 0657-56 is a distant cluster ($z=0.296$), originally detected in the {\it Einstein} survey \citep{tuc95}. ROSAT and ASCA have shown that the Bullet is one
of the hottest (kT=17.4$ \pm 2.5$\,keV), and most massive 
cluster known \citep{tuc98}.
%X-ray data, it was identified as one of the hottest  ($T=17.4 \pm 2.5$ keV), and therefore most massive clusters \citep{tuc98}. 
The same data show that 1E 0657-56 is undergoing a major merger process. 
\citet{lia00} found out that 1E 0657-56 contains a very luminous
radio halo whose surface brightness closely follows the X-ray one.
Weak and strong lensing reconstruction of the Bullet Cluster are one of the best
evidence for the existence of dark matter \citep[e.g.][]{ clo04,mar04,clo06,bra06}.
The {\it Chandra} high resolution image of a bullet-like gas cloud moving in the cluster core with a  bow shock front, gained 1E 0657-56   the name Bullet Cluster  \citep{mar02}. The average temperature they report ranges from 14-15 keV to more 
than 20 keV. Deeper {\it Chandra} observations showed that, away from the bullet,
the radio halo peak is offset from the X-ray peak, which is centered on region hosting the hottest gas  \citep{gov04}. The Bullet Cluster was observed also with  
XMM-Newton \citep{zha04,fin05,zha06}
and  RXTE \citep{pet06}. The latter determined that the spectrum of the
Bullet cluster can be fit equally well by the sum of two thermal
models or by the sum of a thermal and a power-law model.
They also estimated that the equipartition value of the magnetic
field intensity is $\sim$1.2\,$\mu$G. 
1E 0657-56 has a complex radio morphology. The diffuse radio halo 
detected by \citet{lia00}   has a flux density $S_R=$78\,mJ at 1.3\,GHz with a 
spectral index is $\alpha=1.2$.

\begin{figure*}[ht!]
  \begin{center}
  \begin{tabular}{cc}
    \includegraphics[scale=0.4]{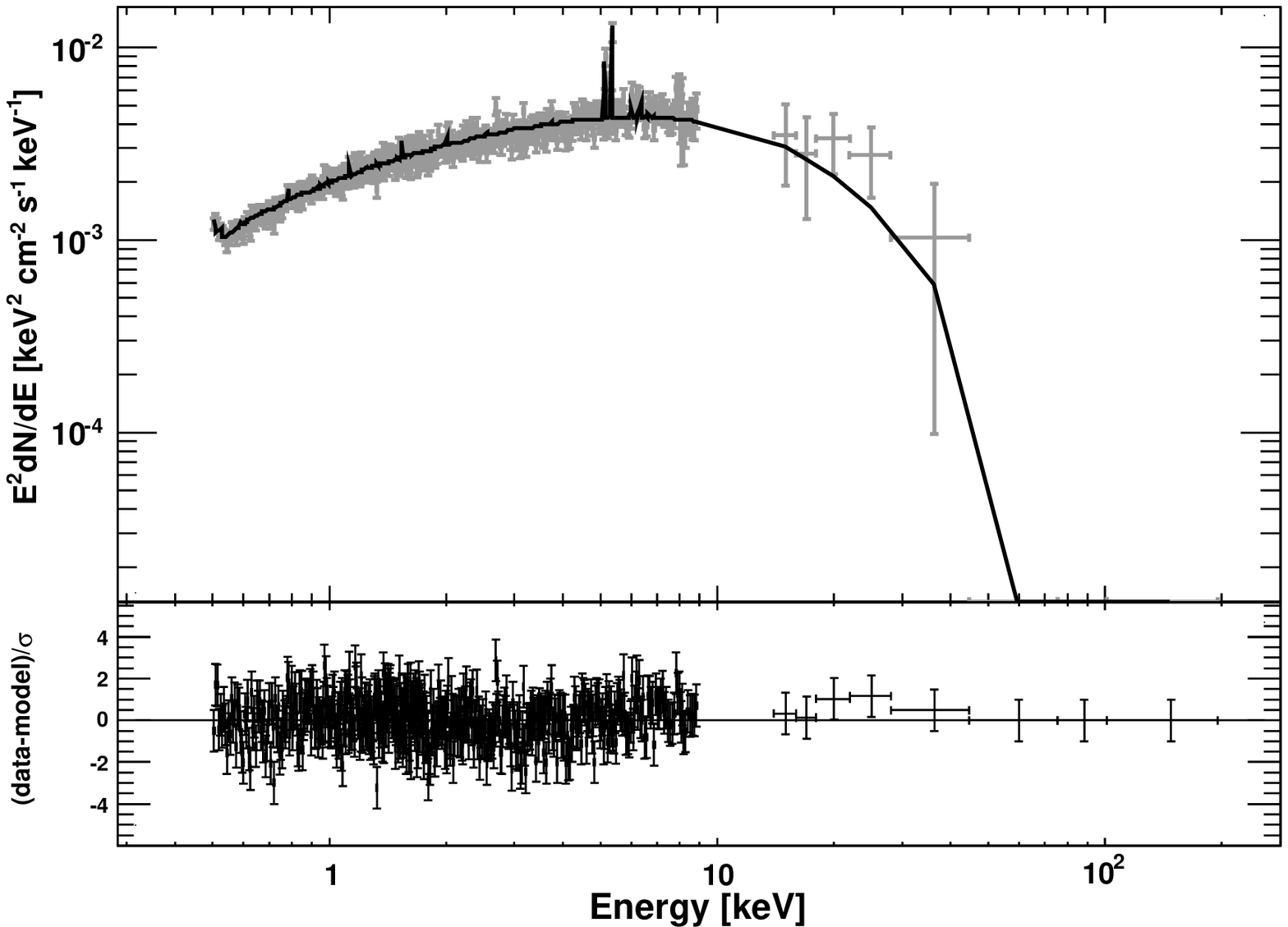}&  %{bullet_singleT.eps} &
	    \includegraphics[scale=0.4]{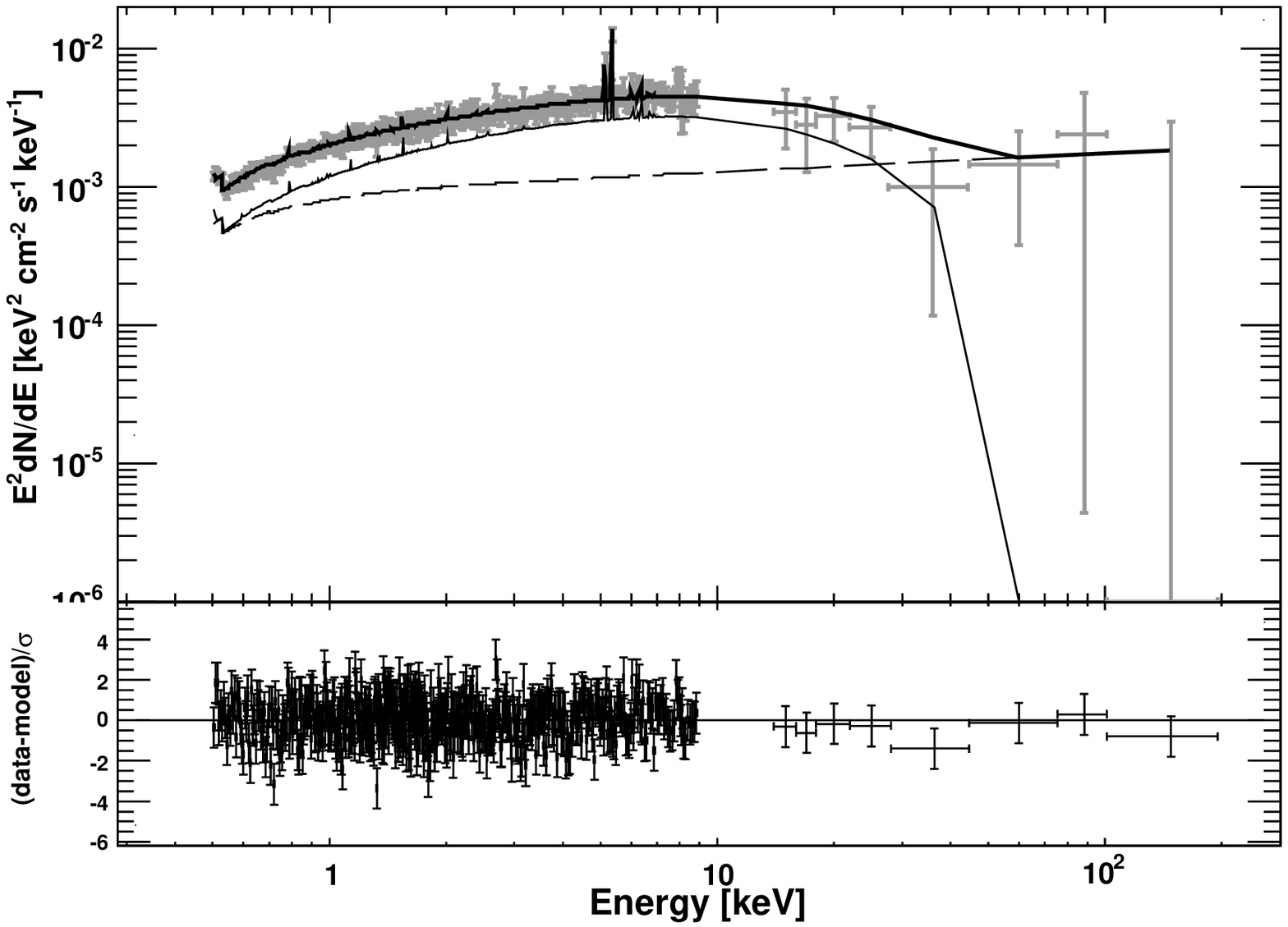}\\%{bullet_pow.eps} \\
 \includegraphics[scale=0.4]{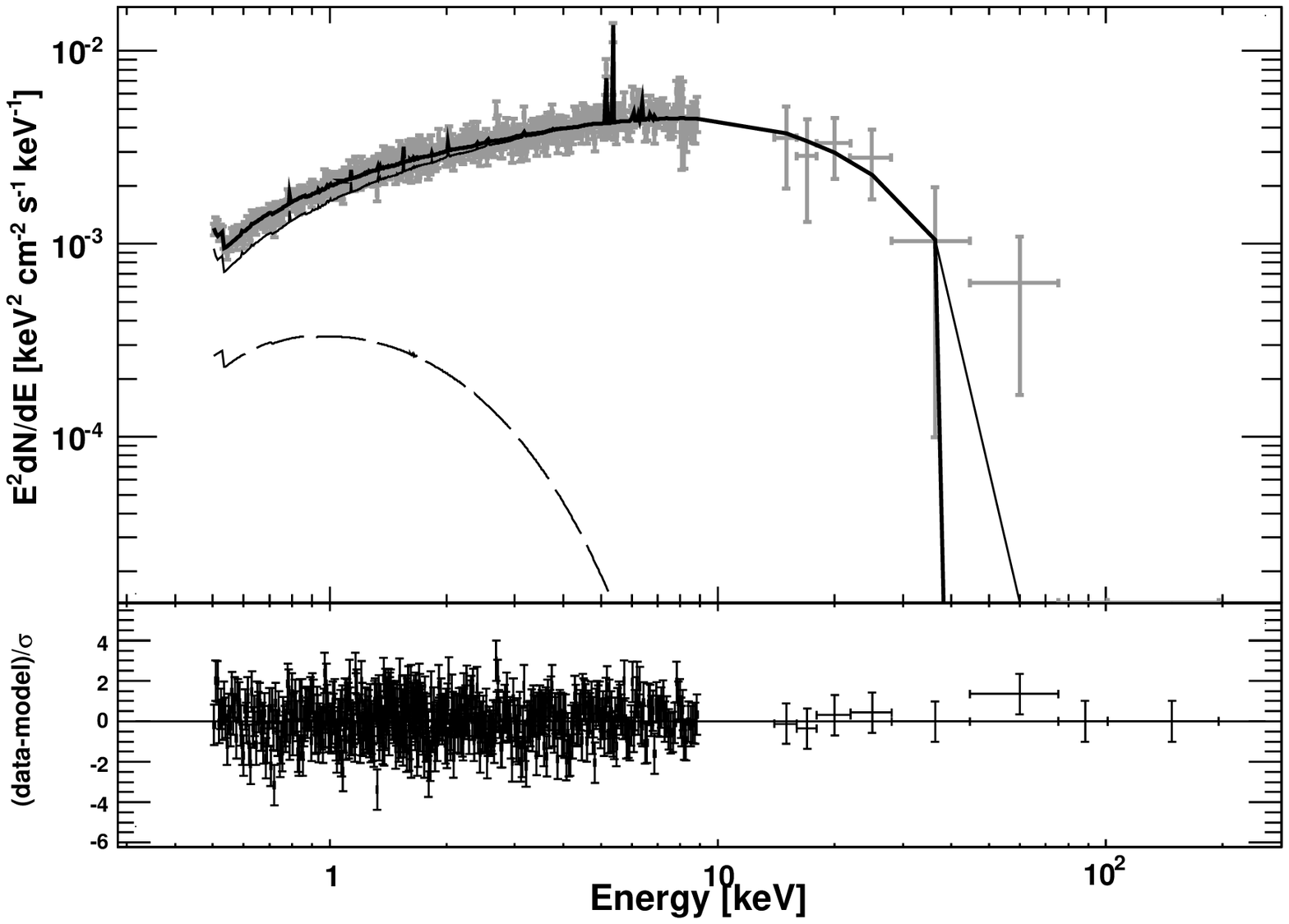}&     %{bullet_2T.eps} &
 \includegraphics[scale=0.4]{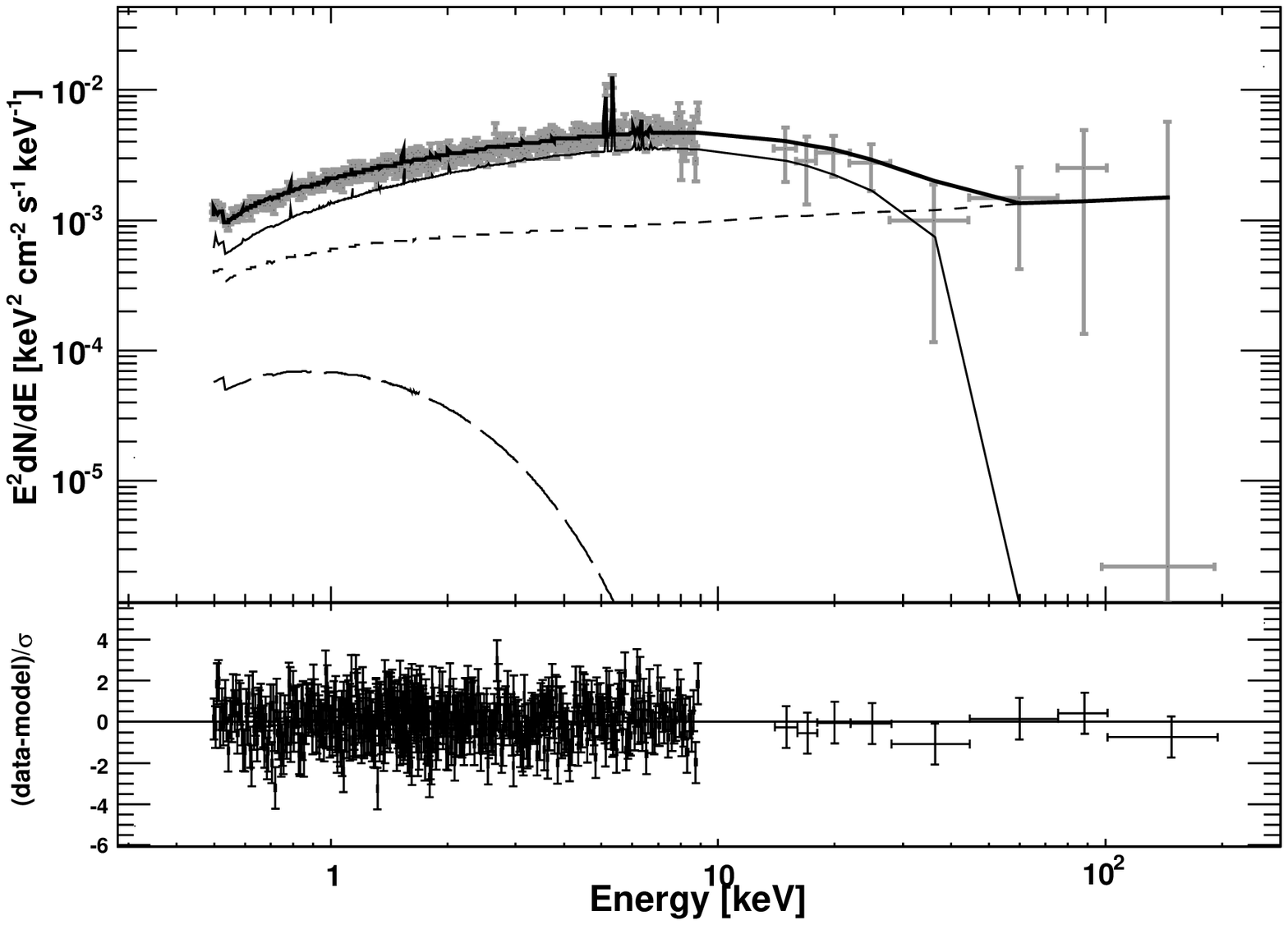}\\    %{bullet_2T_pow.eps}\\

\end{tabular}
  \end{center}
  \caption{
XMM-Newton and BAT data for the Bullet cluster fitted
with: 1) a single thermal model (upper left),
2) the sum of a thermal and a power-law model (upper right), 
3) the sum of two thermal models (bottom left), and
4) sum of two thermal models (thin and long-dashed lines)
 and a power law (bottom right).
}
  \label{fig:bullet}
\end{figure*}

The combined XMM-Newton and BAT data can be successfully fit ($\chi^2/dof$=524.7/513) be a 
single-temperature thermal (APEC) model (see left panel of Fig.~\ref{fig:bullet}). The best-fit temperature
is 12.57$^{+0.64}_{-0.65}$\,keV while the abundance is 0.25$^{+0.06}_{-0.08}$ 
solar. The temperature is in moderate good agreement with the 
values of 14.5$^{+2.0}_{-1.7}$\,keV and 14.8$^{+1.7}_{-1.2}$\,keV
as observed by ASCA and {\it Chandra} respectively \citep{liang00,mar02} and
is well contained in the range of temperatures observed with {\it Chandra} 
(see  above references).
Following \cite{pet06} we added a power-law  model to the fit.
The fit improves ($\Delta \chi^2\approx$23 for 2 additional
parameters, i.e. 4.4\,$\sigma$) and the 
best-fit temperature becomes 14.77$^{+1.13}_{-0.72}$\,keV (in agreement
with ASCA and {\it Chandra} results),  while
the power-law index is 1.86$^{+1.25}_{-0.14}$ . This model fit
is reported in the right panel of Fig.~\ref{fig:bullet}.
The non-thermal 20--100\,keV flux is 3.4$^{+1.1}_{-1.0}\times10^{-12}$\,erg
cm$^{-2}$ s$^{-1}$. 
These values are in  good agreement with those reported by \cite{pet06}.
Moreover, recently \cite{million09}, using Chandra,
reported the detection of non-thermal flux in 
the 0.6--7.0\,keV band at a level of 
0.95$^{+0.10}_{-0.11}\times10^{-12}$\,erg cm$^{-2}$ s$^{-1}$.
From our analysis we derive that the 0.6--7.0\,keV power-law flux is 
3.3$^{+1.2}_{-1.1}\times10^{-12}$\,erg cm$^{-2}$ s$^{-1}$ and thus
slightly brighter than their reported flux.

For the sake of completeness we also tried to fit the spectrum
of the Bullet cluster with the sum of two thermal models.
The hot, most intense component, shows as before a temperature
of 15.4$^{+2.4}_{-1.5}$\,keV and an abundance of 0.30$^{+0.10}_{-0.08}$ solar.
The second component displays a temperature of 1.1$^{+0.4}_{-0.2}$\,keV,
 an abundance compatible with zero and a 1--10\,keV flux of 
6.3$^{+3.8}_{-2.1}\times10^{-13}$\,erg cm$^{-2}$ s$^{-1}$.
This fit is shown in the lower panel of  Fig.~\ref{fig:bullet}.
Both fits represents a reasonable description of the data.
The thermal plus power-law model is slightly worst ($\chi^2/dof$=501.7/511)
than the sum of two thermal models ($\chi^2/dof$=499.9/510). However inspection of Fig.~\ref{fig:bullet}
shows that the thermal plus power-law model explains better the residuals
at high-energy, albeit the BAT data are not very significant above 50\,keV.
We also checked that leaving the inter-calibration
between BAT and XMM-Newton free to vary (see $\S$~\ref{sec:xmmbat}) does not
changes the results presented here.

Finally we also verified the contribution of point sources to the
overall signal. We found that only two sources produce a signal
comparable to the excesses seen here. The first one is a bright
point sources locate south-west of the cluster core at
a position R.A.(J2000)=06:58:13.8 and Decl.(J2000)=-55:59:20.6.
Its spectrum can be fit by an absorbed bremsstrahlung model, where
the absorption is compatible with the Galactic one and the
temperature of the plasma is 3.9$^{+13.1}_{-2.31}$\,keV.
The 1--10\,keV flux is 2.90$^{+1.01}_{-1.50}\times10^{-14}$\,erg cm$^{-2}$ s$^{-1}$.  The second one is located at 
R.A.(J2000)=06:58:03.8 and Decl.(J2000)=-56:01:13.1 and its spectrum
can be fit with an absorbed power law where the absorption
is in excess of the Galactic one with
N$_{\rm H}=$9.3$^{+7.7}_{-4.4}\times10^{21}$\,cm$^{-2}$ and the
photon index is 1.46$^{+0.55}_{-0.31}$. When extrapolated to the 20--100\,keV
band the source flux is 5.0$^{+1.0}_{-2.45}\times 10^{-13}$\,erg 
cm$^{-2}$ s$^{-1}$. It is thus clear that both sources cannot account
for the observed signals, indeed their fluxes are a factor $\sim$10 below
the flux of the 'cold' component
and the non-thermal component seen in the spectral fits described above.

Moreover, we checked whether the results reported above
might be connected to some residual background contamination in XMM-Newton
which was not accounted for correctly in the extraction of the
background spectrum from blank field observations. For this purpose,
we extracted a background spectrum from a region of the XMM-Newton
CCD which is the least contaminated by the cluster emission.
Since the Bullet cluster is at moderately high-redshift this is possible.
We also  extracted the background spectrum from
a region whose area was the same as that one used to extract the cluster
emission and  whose position is 
diametrally opposed to the cluster region with respect to the pointing
direction (to ensure a similar effective area over the two regions).
The analysis of the Bullet cluster spectrum using this background strategy
confirms the above results. In particular both the 'cold' component
and the non-thermal power law are confirmed and the derived fluxes
are consistent with those reported above. Thus, we can exclude an instrumental
origin for both components.

In order to understand whether both components co-exist, we tried
a fit with a model which is the sum of two thermal component and 
a power law. Not surprisingly this model produces a good fit to the data
($\chi^2/ndf$=498.1/508). All parameters, with the exception of the
normalization of the 'cold' component, are fully compatible and in
good agreement with the parameters reported above.
Indeed, from this best-fit we derive that the cold component has now
a 1--10\,keV flux of 
7.7$^{+1.1}_{-7.7}\pm10^{-14}$\,erg cm$^{-2}$ s$^{-1}$ and thus
a factor of $\sim10$ below the flux derived from the fit using two thermal
models only. Moreover, as the statistical uncertainty shows, this component
is now compatible with zero at 90\,\% confidence.

\begin{figure*}[ht!]
  \begin{center}
  \begin{tabular}{c}
    \includegraphics[scale=0.33,angle=270]{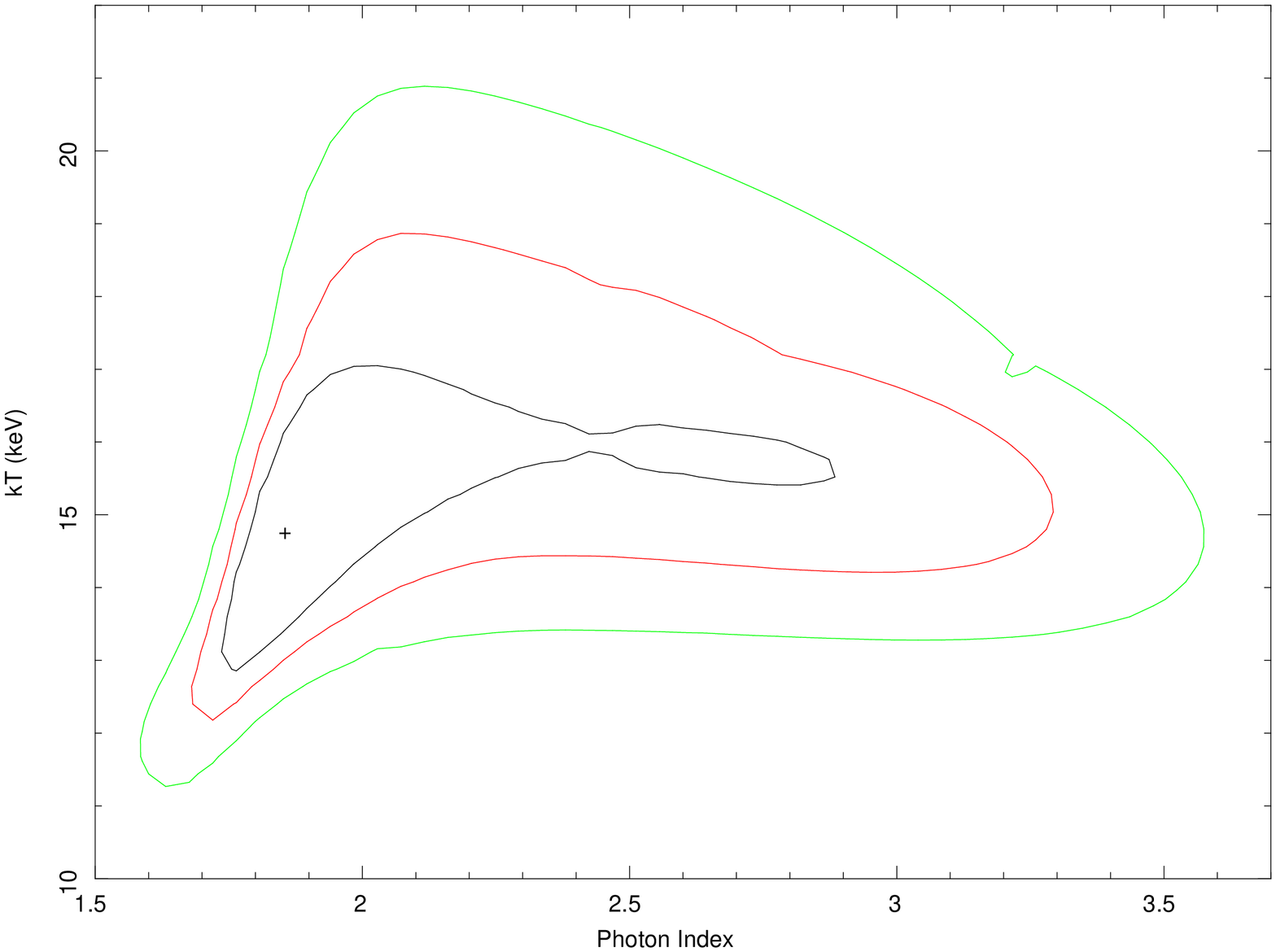}%{cont_kTgamma.eps} 
 \includegraphics[scale=0.33,angle=270]{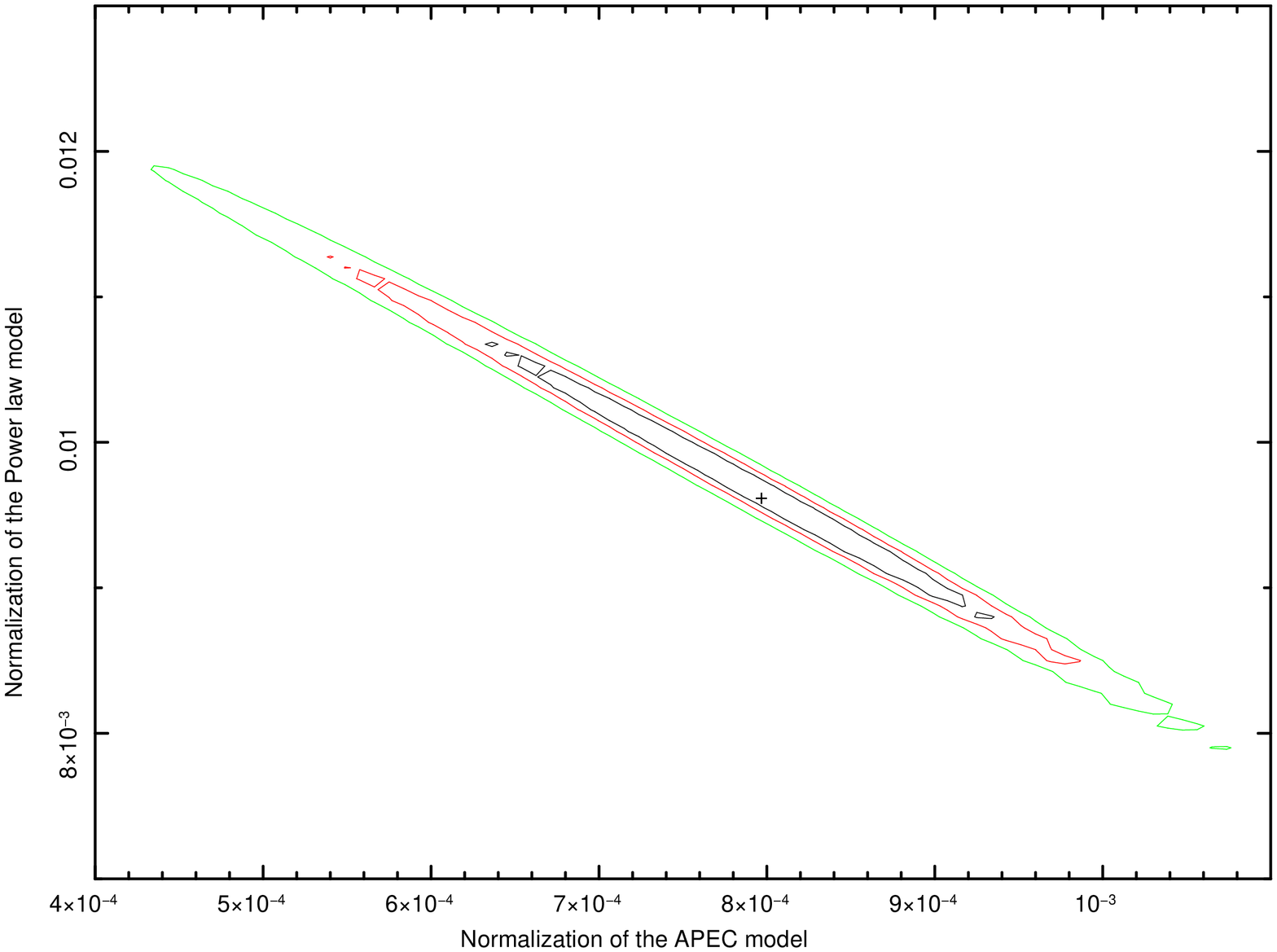}\\%{cont_norm.eps}\\
\end{tabular}
  \end{center}
  \caption{Contour (1,2, and 3\,$\sigma$)
plots of the parameters of the power-law component
versus the parameters of the most intense thermal component
for the Bullet cluster.
}
  \label{fig:cont}
\end{figure*}

On the other hand, the parameters
of the power-law component are robust with respect to the variation of the
other parameters (see Fig.~\ref{fig:cont} ). For this and all the other reasons explained above
we believe that
the description of the cluster spectrum in terms of a single thermal
model and a power law is the best and most reliable one.
The parameters of this fit are reported in Tab.~\ref{tab:spec}.
Our analysis thus confirms the presence of a power-law component
in the spectrum of the Bullet cluster as reported by \cite{pet06}.

%%%%%%%%%%%%%%%%%%%%%%%%%%%%%%%%%%%%%%%%%%%%%%%%%%%%%%%%%%%%%%%%%%%%
  \subsection{PKS 0745-19}

Early  {\it Einstein} and EXOSAT observations showed that  PKS 0745-19
(also known as 4U 0739-19) is one of the largest cool core
cluster known \citep{fab85,arn87,edg90,whi97}.
The ICM temperature as measured with ASCA and ROSAT \citep{all96,pie98,per98}
agrees well with that one found with BeppoSAX, of about 8.3\,keV
\citep{deg99,deg01,deg02}.
\cite{che03}, using XMM-Newton and {\it Chandra}, reported an enhanced 
diffuse X-ray
emission in correspondence of  bright radio lobes.
%They exclude that such excess is due to Inverse Compton  on CMB photons
They discuss the possibility  of  buoyant bubbles to explain the observed
 X-ray  and radio emission. 
However \cite{dun06}, using {\it Chandra} data, find
"no clear" evidence of such radio bubbles. 
\citet{bal93} and \cite{bau91} classify PKS 0745-19 as an amorphous 
radio source, displaying both
a compact radio source (five times brighter than Perseus) and a 
diffuse emission. These unusual radio properties do not seem just the 
result of AGN activity, but the result of a merger and/or buoyant 
plumes.  \cite{bau91} report that 
the intensity of  the diffuse flux at 5\,GHz is 265\,mJy, 
while the spectral index  is $\alpha=-1.4$.
For these values, 
the equipartition magnetic field is  in the   20--50\,$\mu$G   range.

\begin{figure*}[ht!]
  \begin{center}
  \begin{tabular}{cc}
    \includegraphics[scale=0.4]{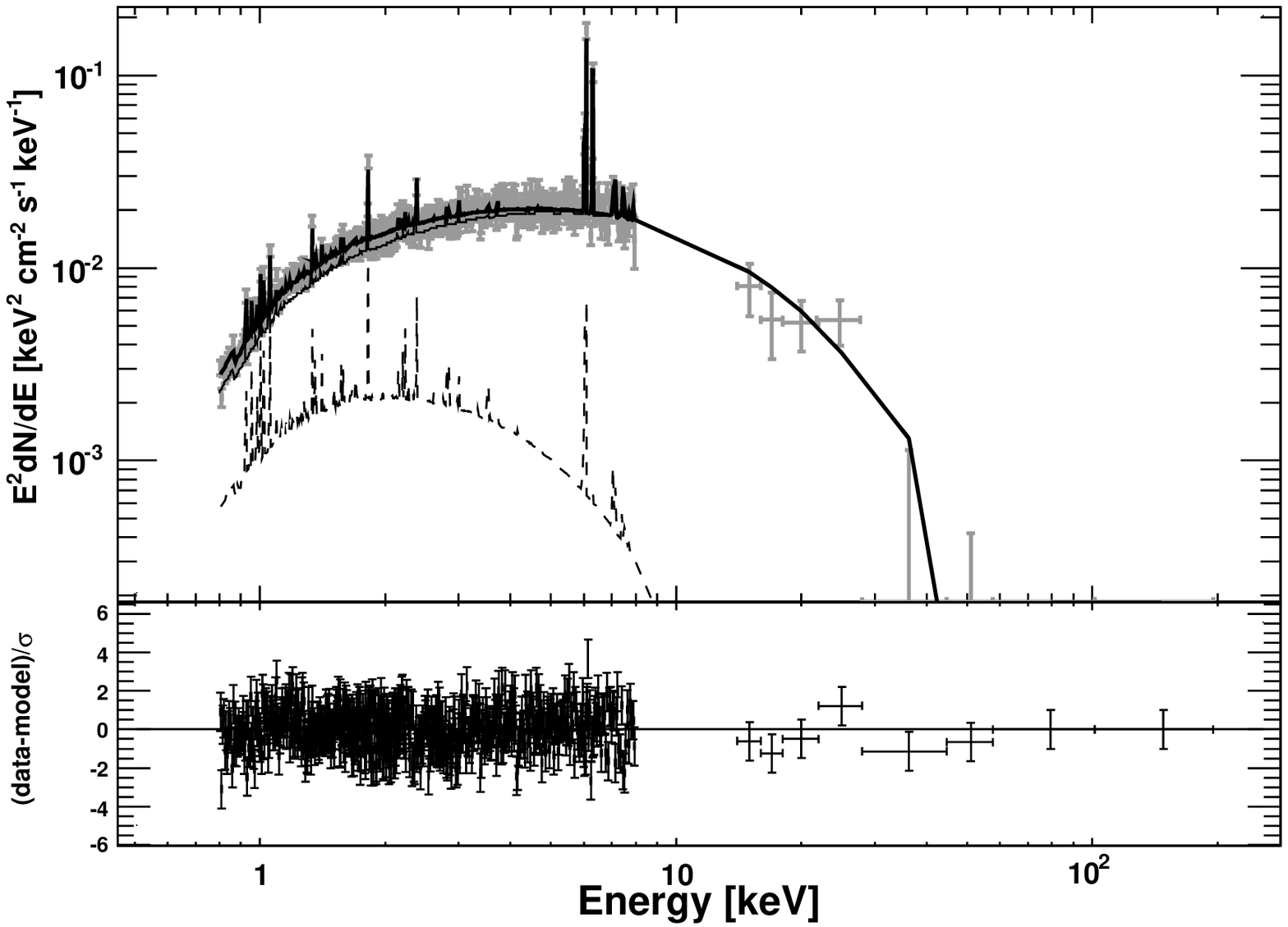}%{pks0745_2T.eps} 
 \includegraphics[scale=0.4]{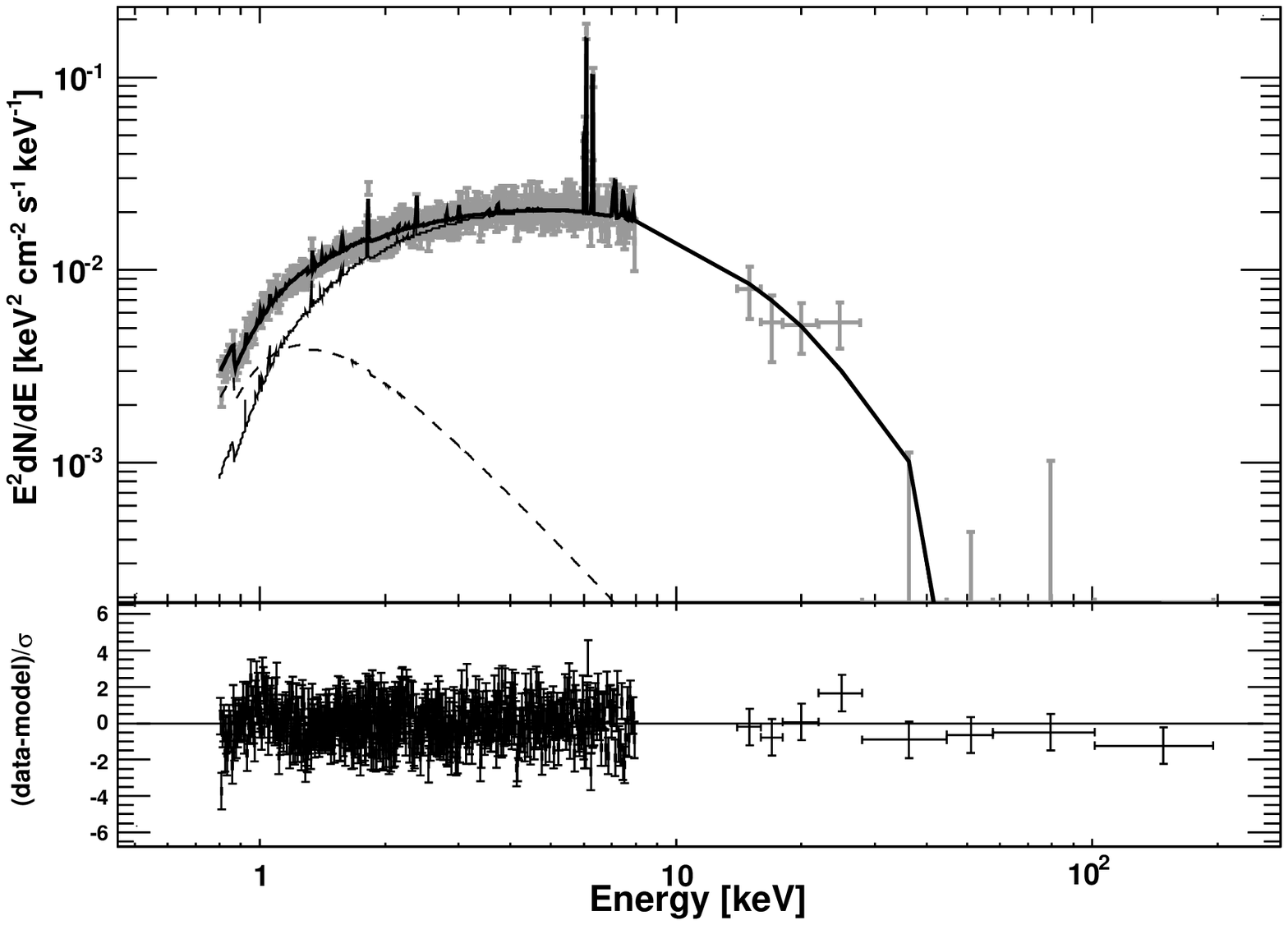}%{pks0745_pow.eps} 

\end{tabular}
  \end{center}
  \caption{{\bf Left Panel:}
Spectrum of PKS 0745-19 fitted with the sum of two thermal models.
{\bf Right Panel:} Spectrum of PKS 0745-19 fitted with
the sum of a thermal model and a power law (dashed line).
}
  \label{fig:pks0745}
\end{figure*}

Our data point to a thermal origin of the hard X-ray emission.
Indeed, the XMM-Newton and BAT spectra can be successfully modeled by 
a single temperature thermal model ($\chi^2/dof$=610.5/581) with
a temperature of 6.69$^{+0.25}_{-0.27}$\,keV and an abundance of
0.35$^{+0.03}_{-0.03}$. However, adding a second thermal component
produces a noticeable improvement ($\Delta\chi^2=23.2$ for 3 additional
degrees of freedom, corresponding to 5.0$\times10^{-5}$ chance).
The hot most intense component has now a temperature of 
7.96$^{+0.68}_{-0.54}$\,keV and an abundance of 0.40$^{+0.05}_{-0.05}$,
while the cold component displays a temperature of 2.16$^{+1.08}_{-0.56}$\,keV
and an abundance of 0.31$^{+0.05}_{-0.06}$. This fit is shown in the 
left panel of Fig.~\ref{fig:pks0745}. The temperatures observed
here are consistent with those found by \cite{george09} using {\it Suzaku}.

We also tried a fit with the sum of a thermal and a power-law model.
The temperature and abundance of the thermal component are
7.53$^{+0.27}_{-0.69}$\,keV and an abundance of 0.37$^{+0.10}_{-0.07}$,
while the photon index of the power law is 4.39$^{+0.79}_{-2.01}$.
The improvement in the $\chi^2$ with respect to the single temperature
thermal model is $\Delta \chi^2=17$ for 2 additional parameters 
(corresponding to 0.2\,\% chance). This fit is reported in the right
panel of Fig.~\ref{fig:pks0745}. Nevertheless, given the very soft
spectral index (which accounts for the 'cold' component) and the
marginal improvement in the fit statistics, we believe that the
sum of two thermal models is a more reliable interpretation of the
XMM-Newton/BAT dataset. The parameters of this fit are thus
summarized in Tab.~\ref{tab:spec}.

Using the sum of two thermal models as a baseline spectral fit
we estimated the 99\,\% CL to a non-thermal component in the 
50--100\,keV band using a power law with a photon index of 2.0.
This is found to be 1.55$\times10^{-12}$\,erg cm$^{-2}$ s$^{-1}$.
If instead a photon index of 2.4 is used, the upper limit
in the 50--100\,keV band becomes 1.10$\times10^{-12}$\,erg cm$^{-2}$ s$^{-1}$.
Finally, the brightest point-source in the XMM-Newton 10\arcmin\ 
region is located at R.A.(J2000) = 07:47:19.0 Decl.(J2000) = -19:24:02.3. 
Its spectrum is very
soft and consistent with a bremsstrahlung model with a temperature
of 0.7$\pm0.3$\,keV. Its flux in the 2--10\,keV band is 2.6$\times10^{-15}$\,erg
cm$^{-2}$ s$^{-1}$ and thus it is negligible when compared to the cluster
signal in both the XMM-Newton and BAT bands.

%%%%%%%%%%%%%%%%%%%%%%%%%%%%%%%%%%%%%%%%%%%%%%%%%%%%%%%%%
%%%%%%%%%%%%%%%%%%%%%%%%%%%%%%%%%%%%%%%%%%%%%%%%%%%%%%%%% Abell 1795
\subsection{Abell 1795}

Abell 1795 is a compact and rich cluster with a strong cool core 
\citep[e.g., ][]{edg92,bri96,mar98,tam01}.
It has been extensively observed with  HEAO-1, {\it Einstein}, EXOSAT,
ROSAT, BeppoSAX, XMM-Newton, {\it Chandra} and RXTE \citep[e.g. ][ and 
references therein]{kow84,rhe91,edg90,arn91,mar98,san00,
arn01,nev04,rev04,vik05}.
EGRET was also used to set an upper limit to the 
$\gamma$-ray  emission from the ICM \citep[above 100 MeV, ][]{rei03}. 
Outside the cool region, the average temperature is of 6--7\,keV 
\citep[e.g.][]{arn01}. Abell 1795 is a relaxed cluster, although 
there is evidence that the central brightest galaxy is not at rest 
\citep{hil88} and that there is inner gas sloshing in the potential 
well \citep{mar01,ett02}.
 No hard X-ray excess has ever been reported for this cluster. 
\cite{fab01} discovered a 
40\arcsec\ long cold filament in the core of the cluster, 
in coincidence with an H$\alpha$ filament. The straightness of the filament 
indicates that the ICM is not very turbulent. Abell 1795 has two X-ray dim 
regions \citep[e.g., ][]{fab01,ett02} in correspondence of radio bubbles 
\citep{dun05}. The radio morphology  is dominated  by two 
radio regions \citep{gut74,owe75,dag82, ali83, bur90, owe93,owe97,dun05}, 
but there is no strong evidence for a  
large scale radio halo  \citep{han82}.
\cite{ge93} report  Faraday rotation measurements of the small central 
radio galaxy 1346+268. They conclude that
 the associated magnetic field must be $>$20\,$\mu$G and that it is
 most likely  associated with the ICM rather than with the small radio source.

\begin{figure*}[ht!]
  \begin{center}
  \begin{tabular}{cc}
    \includegraphics[scale=0.4]{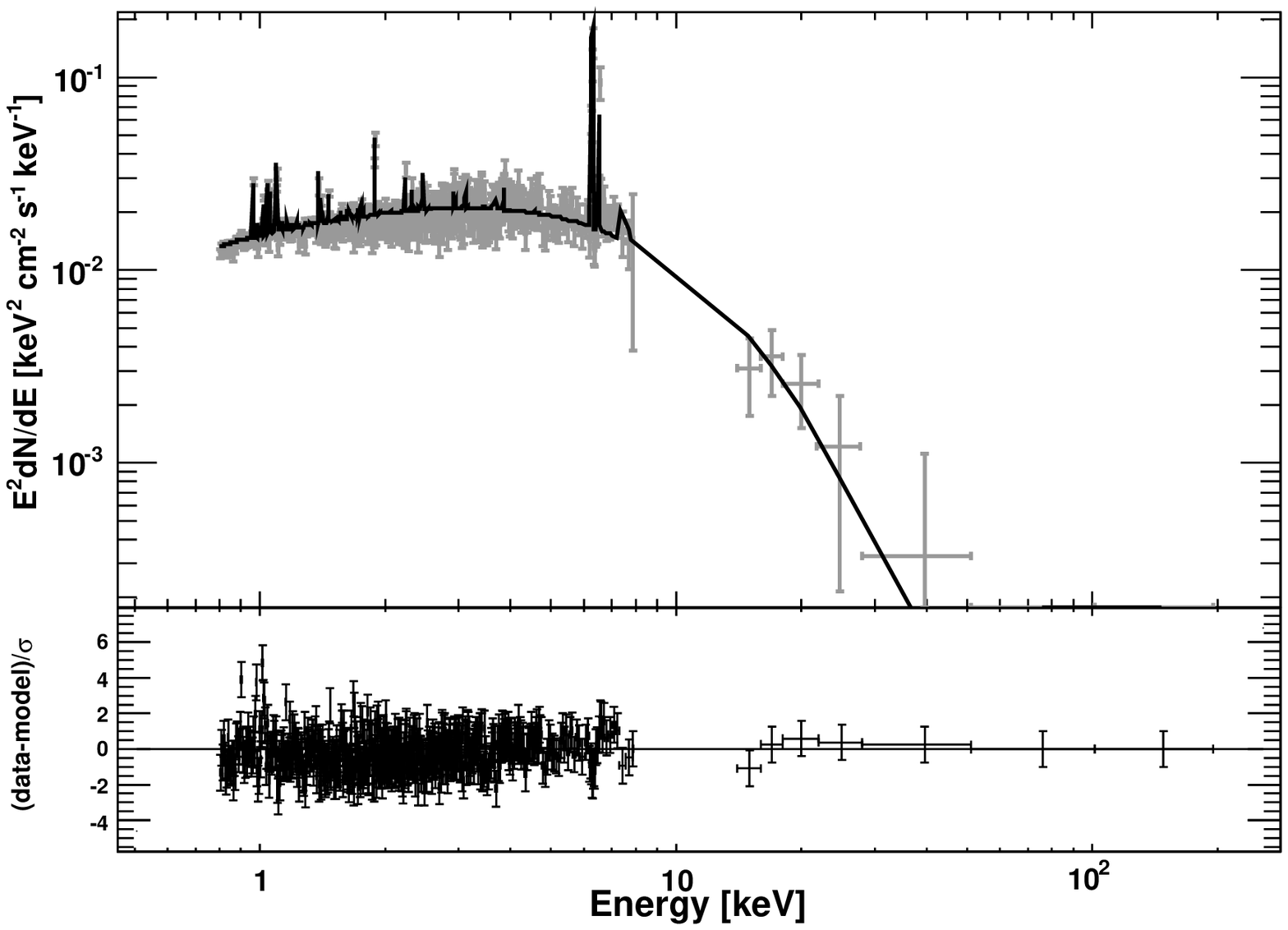}%{abell1795.eps} 
	 \includegraphics[scale=0.4]{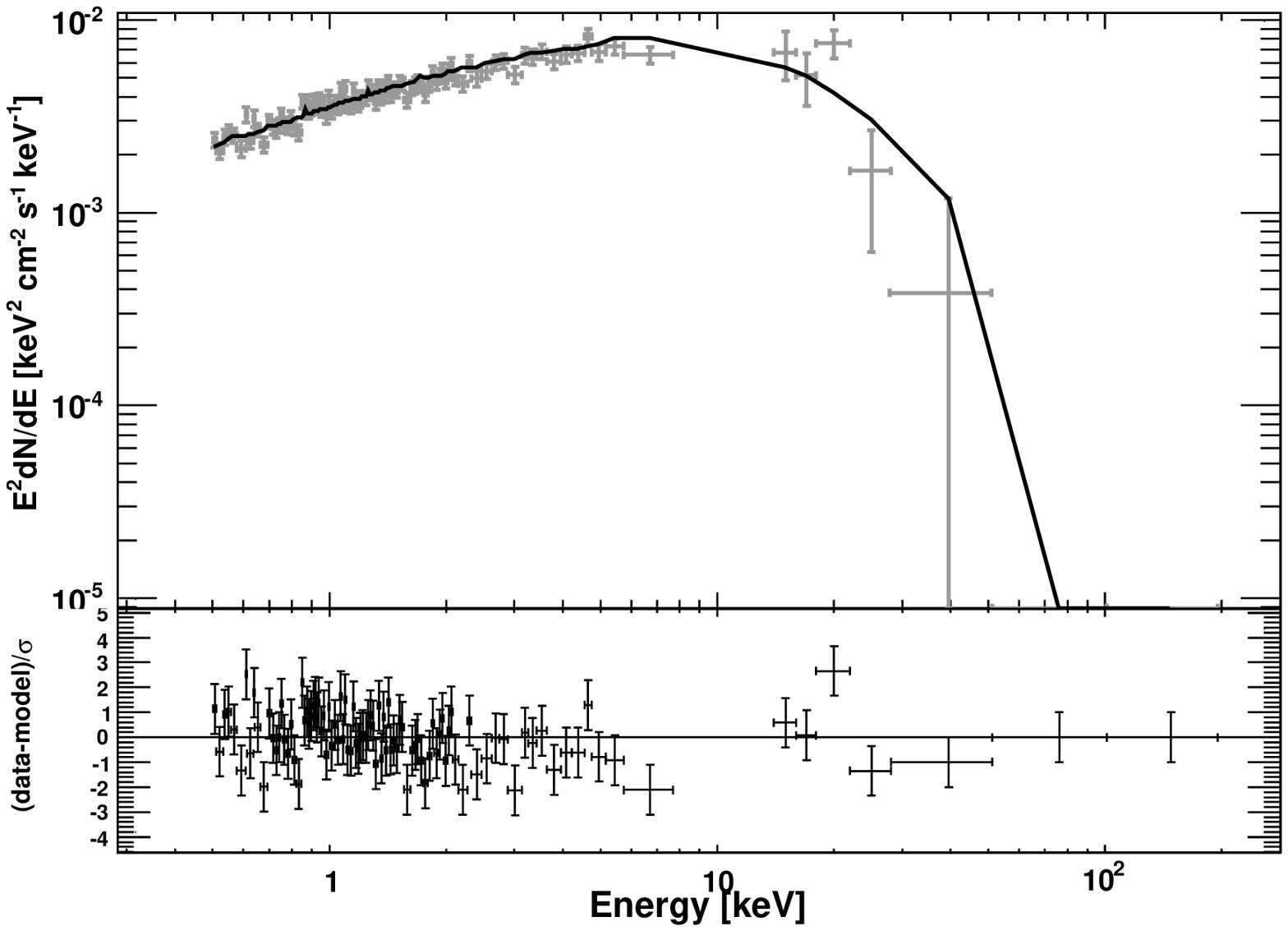}\\%{a1914spec.eps}\\
\end{tabular}
  \end{center}
  \caption{{\bf Left Panel:}
Spectrum of Abell 1795 fitted with a single temperature thermal model.
{\bf Right Panel:}
Spectrum of Abell 1914 fitted with a single temperature thermal model.
}
  \label{fig:a1795}
\end{figure*}

Using the parameters of the surface brightness reported by \cite{bri96}
(e.g. core radius of 5.15\arcmin\ and $\beta=$0.93), we derive
that our standard selection region of 10\arcmin\ radius (in XMM-Newton)
includes 97--99\,\% of the cluster emission.
The XMM-Newton and {\it Swift}/BAT data are best  modeled
($\chi^2/dof$ = 892.1/1275) by a single thermal model
with a temperature of 4.82$^{+0.10}_{-0.11}$\,keV
and an abundance of 0.45$^{+0.04}_{-0.04}$. 
Our results are in good agreement with the ones of \cite{tam01}.
Given the very good $\chi^2$, adding other models (or free parameters)
does not improve the fit. Thus, we believe that the single thermal
model is a good representation of the XMM-Newton/BAT dataset.
This fit is shown in the left panel of Fig.~\ref{fig:a1795}.

As a final note, the field of Abell 1795 contains a bright AGN
located at R.A.(J2000)=13:48:35.2 and Decl.(J2000)=26:31:08.9,
which lies less than 3\arcmin\ from the cluster core.
This source is  the Seyfert 1 galaxy 1E 1346+26.7.
In XMM-Newton it exhibits an unabsorbed  power-law spectrum
with a photon index of 2.34$\pm0.04$ and a 2--10\,keV flux of
8.7$(\pm0.1)\times10^{-13}$\,erg cm$^{-2}$ s$^{-1}$ (approximately 1\,\%
of the cluster's signal in that band).
When extrapolated to the 15--55\,keV band, the AGN flux becomes 
3.7$(\pm0.1)\times10^{-13}$\,erg cm$^{-2}$ s$^{-1}$ and thus
still a factor $\sim5$ below the cluster emission in that band 
(see Tab.~\ref{tab:spec}).
We checked that the results reported above do not change if
the AGN emission is properly modeled (e.g. with the parameters
of the AGN power-law allowed to vary within their errors) in the
joint fit to the overall cluster's spectrum.
However, when computing the (50--100\,keV) 99\,\% CL
upper limit on the non-thermal
component (using a power-law with a photon index of 2.0)
we derive that this is 1.57$\times10^{-12}$\,erg cm$^{-2}$ s$^{-1}$
if the AGN contribution is not taken into account, or
1.06$\times10^{-12}$\,erg cm$^{-2}$ s$^{-1}$ if it is.
We will thus use this second upper limit in Tab.~\ref{tab:ulxmm}.

%%%%%%%%%%%%%%%%%%%%%%%%%%%%%%%%%%%%%%%%%%%%%%%%%%%%%%%%%%%%%%%%%%%%
\subsection{Abell 1914 }

Abell 1914 is a regular and smooth galaxy cluster. It has 
been observed with ROSAT \citep{ebe96,buo96,boh00}, ASCA 
\citep{whi00,ike02} and {\it Chandra} \citep{gov04,bal07}. 
By means of a comparison of X-ray and radio maps,
 \cite{gov04} discuss a possible merger scenario.
 % [OTTIMA DESCRIZIONE] in gov04 
\cite{bal07} report an   average ICM temperature of $9.20\pm0.39$\,keV.
Abell 1914 is known to host a very steep radio source \citep{kul90}
and a radio halo  \citep{gio99,kem01,bac03}. 
The point sources make the estimate of the 
diffuse flux density difficult.
We adopt the value of  $S_R$=64\,mJ at 1.4\,GHz and
the reported spectral index of $\alpha=1.8$ from \cite{bac03}.  
The equipartition  magnetic field is 0.5\,$\mu$G.

The {\it Swift}/BAT spectrum can be fit by a bremsstrahlung model
with a plasma temperature of 7.30$^{+3.18}_{-2.01}$\,keV.
The combined XMM-Newton -- {\it Swift}/BAT dataset can be successfully modeled
($\chi^2/dof$=355.1/351)
with a single-temperature thermal model (see  right panel of
Fig.~\ref{fig:a1795}).  
The best-fit temperature
and metallicity are 11.14$^{+1.13}_{-1.09}$\,keV and 0.19$\pm0.14$ solar,
in agreement with the studies mentioned above.
Given the good $\chi^2$ adding other models to the single thermal
model does not improve the fit results.
The 99\,\% CL upper limit  on the 50--100\,keV non-thermal flux,
 evaluated with a power law with a photon index of 2.0,
flux is 1.08$\times10^{-12}$\,erg cm$^{-2}$ s$^{-1}$.
If we use a power law with a photon index of 2.8 (in line with the
radio photon index) then the upper limit is much tighter and it becomes
4.60$\times10^{-14}$\,erg cm$^{-2}$ s$^{-1}$.
However, this upper limit to the non-thermal flux in the (BAT) 50--100\,keV
band, is entirely driven by the XMM-Newton signal below 2\,keV.
Indeed, if we repeat the same process described above, but using only
{\it Swift}/BAT data, then the 99\,\% CL upper limit (using a photon
index of 2.8) is  1.16$\times10^{-12}$\,erg cm$^{-2}$ s$^{-1}$
and thus in line with the one computed using a power-law with a photon
index of 2.0 and  the entire XMM-Newton/BAT dataset.
We thus believe that this (e.g. 1.08$\times10^{-12}$\,erg cm$^{-2}$ s$^{-1}$)
is for the 50--100\,keV band a more reliable
upper limit.

%%%%%%%%%%%%%%%%%%%%%%%%%%%%%%%%%%%%%%%%%%%%%%%%%%%%%%%%%%%%%%%%%%%%%%%%%%
%%%%%%%%%%%%%%%%%%%%%%%%%%%%%%%%%%%%%%%%%%%%%%%%%%%%%%%%%%%%%%%%%%%%%%%%%%
\subsection{Abell 2256}

Abell 2256 is a rich cluster at a redshift of 0.0581, bright 
both at radio and X-ray energies \citep[e.g.][]{bridle76,briel91,henriksen99}.
It has been studied several times at X-rays and the disturbed
morphology of the X-ray temperature map indicates a 
cluster in an advanced merging stage \cite[e.g.][]{molendi00}.
Abell 2256 is one of those clusters for which a claim
of significant detection of non-thermal emission has been reported.
Indeed, \cite{fusco00}, using data from BeppoSAX, reported
the detection of a hard X-ray excess at the 4.6\,$\sigma$ level.
The 20--80\,keV flux of this excess is 
1.2$\times 10^{-11}$\,erg cm$^{-2}$ s$^{-1}$.
\cite{rephaeli03}, using RXTE data, reported the detection of an
hard X-ray excess whose  20-80\,keV flux is 
4.3$^{+5.7}_{-4.0}\times 10^{-12}$\,erg cm$^{-2}$ s$^{-1}$ (errors
are 90\,\% CL) and thus a factor of $\sim$3 fainter than
the one reported by \cite{fusco00}, but marginally consistent with it.
A re-analysis by \cite{fusco05} confirmed the BeppoSAX detection
(at 4.8\,$\sigma$) albeit at a lower 20-80\,keV flux of 
8.9$^{+4.0}_{-3.6}\times 10^{-12}$\,erg cm$^{-2}$ s$^{-1}$.
In the radio band, Abell 2256 displays an extremely complex morphology
consisting of a bright relic and a fainter steep-spectrum radio halo located
in the cluster center \citep{clarke06}. The total flux
density of this radio halo is 100\,mJy at 610\,MHz \citep{rengeling97}
and 103.4\,mJy at 1369\,MHz \citep{clarke06}, while the spectral
index is $\alpha$=1.8.

\begin{figure*}[ht!]
  \begin{center}
  \begin{tabular}{c}
    \includegraphics[scale=0.7]{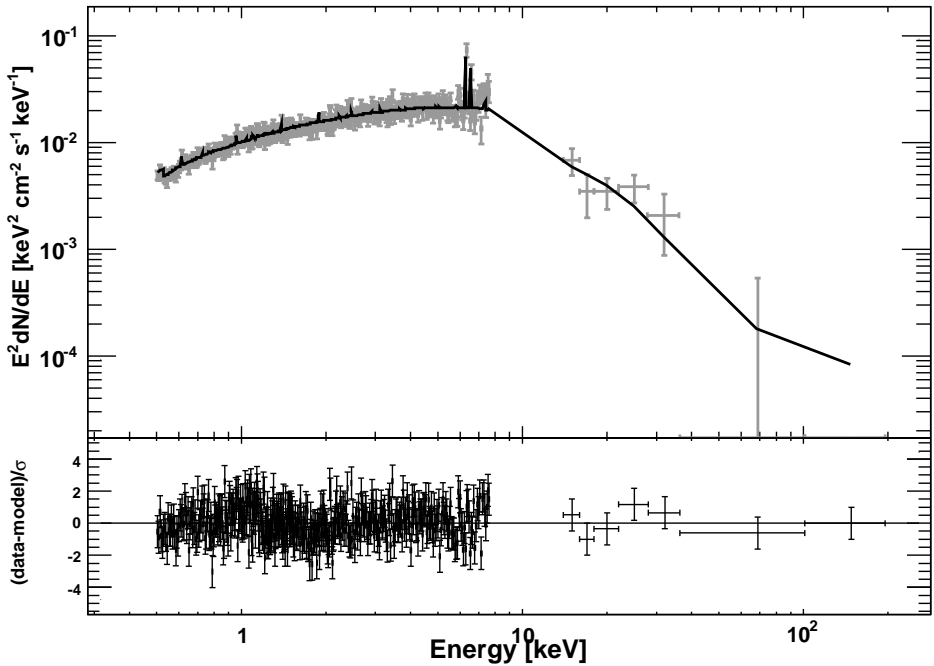}%{abell2256.eps} 
\end{tabular}
  \end{center}
  \caption{
Spectrum of Abell 2256 fitted with a single temperature thermal model.
}
  \label{fig:a2256}
\end{figure*}

The BAT data alone are well fit by a bremsstrahlung
model with a temperature of 9.8$^{+7.7}_{-3.8}$\,keV.
Adopting the values reported by \cite{briel91} (e.g. core radius of 
4.83$\pm0.17$\,arcmin and $\beta$=0.756$\pm0.013$), we derive
that selecting photons within 10\arcmin\ of the core includes $\sim$95\,\%
of the cluster's emission.
The joint XMM-Newton-BAT dataset is well fit ($\chi^2/dof$=445.6/445)
by a single thermal model with a temperature of 
8.84$^{+0.66}_{-0.61}$\,keV and an abundance of 0.22$\pm0.06$.
The best fit is shown in  Fig.~\ref{fig:a2256}
while the parameters are reported in Tab.~\ref{tab:spec}.
\cite{henriksen99} found out that the best spectral model
reproducing the RXTE/ASCA datasets, for Abell 2256, is produced
by the sum of two thermal models. In that work the hot 
and the cold components have a temperature of $\sim$7\,keV  and $\sim$1\,keV
respectively. Following his example, we  added a second thermal
model to the fit keeping the abundance of this additional
component fixed at 0.3 (allowing
this parameter to vary does not change the results).
The best-fit temperature of the additional component is 1.08$\pm0.39$\,keV in good
agreement with the results of \cite{henriksen99} while the
temperature and abundance of the hot component did not vary appreciably.
However, the improvement in the $\Delta \chi^2$ is 2.5 for two additional parameters
and thus not significant (i.e. the probability that the improvement
was obtained by chance is $\sim$0.3). We thus believe that the single
temperature thermal model discussed above represents the best description
of the XMM-Newton-BAT dataset.

The 99\,\% CL upper limit on the 20--80\,keV non-thermal flux
is 6.1$\times10^{-12}$\,erg cm$^{-2}$ s$^{-1}$. The upper limit
derived by our analysis is lower than
the hard X-ray excess claimed by \cite{fusco05}. 
Even using BAT data alone, the 99\,\% upper limit in the 20-80\,keV
band is 4.6$\times10^{-12}$\,erg cm$^{-2}$ s$^{-1}$ and thus
inconsistent with the BeppoSAX result (but not with  the RXTE one).
In our band (50--100\,keV)  the upper limit, derived
from the joint dataset, is 
2.41$\times10^{-12}$\,erg cm$^{-2}$ s$^{-1}$.
If instead of a power law with an index of 2.0, we use a power law
with a photon index of 2.8 (the value of the radio halo) the 50--100\,keV
upper limit on the non-thermal emission would be
 1.97$\times10^{-13}$\,erg cm$^{-2}$ s$^{-1}$.

%%%%%%%%%%%%%%%%%%%%%%%%%%%%%%%%%%%%%%%%%%%%%%%%%%%%%%%%%%%%%%%%%%%%%%%%%%
%%%%%%%%%%%%%%%%%%%%%%%%%%%%%%%%%%%%%%%%%%%%%%%%%%%%%%%%%%%%%%%%%%%%%%%%%%
\subsection{Abell 3627}
\label{sec:norma}

Abell 3627, 
known also as Norma cluster at z=0.015, 
is a nearby massive cluster located behind the Milky Way in the core of the
Great Attractor and discovered as an important component of the local
large-scale structure by  \cite{kraan96}. It  is a very   
rich cluster with a mass comparable to that  of  Coma and Perseus
 (i.e. $>$2$\times$10$^{15}$ M$_{\odot}$). 
Early X-ray observations with ROSAT and ASCA show that
the cluster is not spherically symmetric and has a  strong temperature
gradient ($\Delta kT\sim$3\,keV) in the direction of the elongation
\citep{boehringer96,tamura98}. This fact indicates that Abell 3627
is in the stage of a major merger. This cluster also exhibits spectacular
head-tail radio-galaxies \citep[e.g.][]{sun10} which are galaxies likely 
traveling at high velocities through the ICM \citep[see e.g.][]{sar88}.
At the cluster center, PKS 1610-608 displays, in radio, two powerful jets
and two lobes
whose surface brightness peaks respectively at 
$\sim$1\arcmin\ and $\sim$5\arcmin\
away from the galaxy. \cite{jones} find that the intensity of the 
magnetic field, derived assuming equipartition, is $\sim$15\,$\mu$G
at the position of the jets and $\sim5$\,$\mu$G at the lobes.

\begin{figure*}[ht!]
  \begin{center}
  \begin{tabular}{c}
    \includegraphics[scale=0.7]{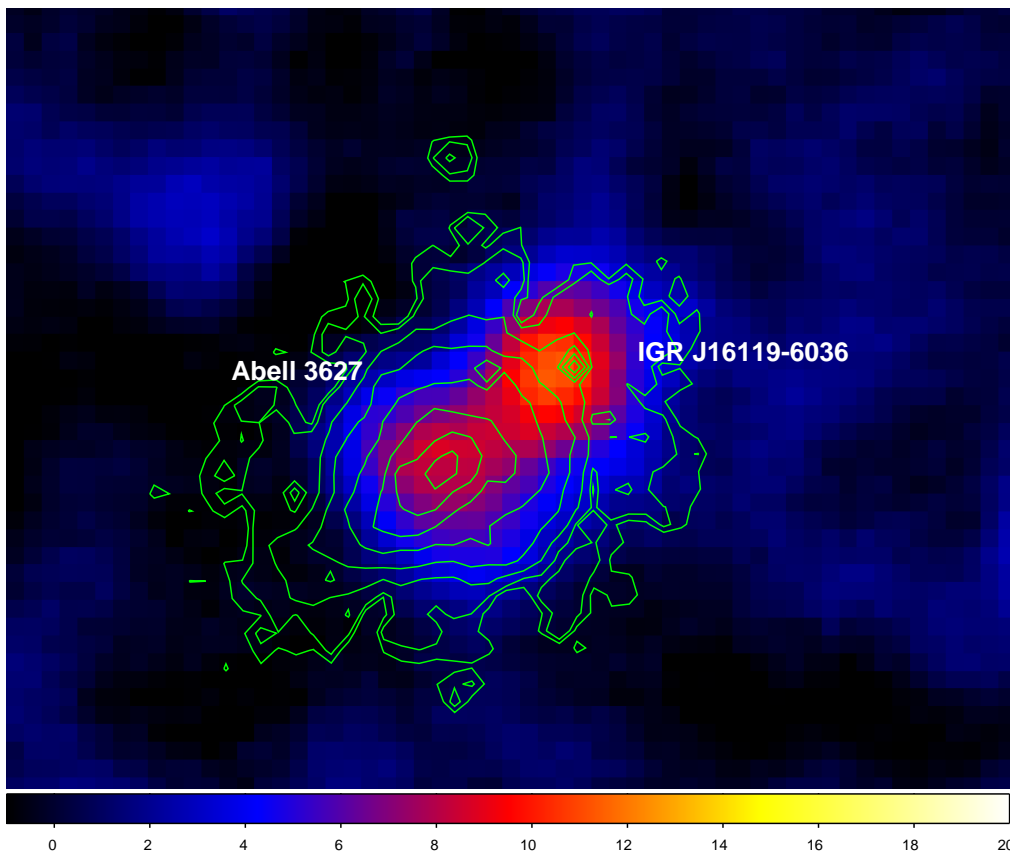}%{a3627_rosatlev.ps} 
\end{tabular}
  \end{center}
  \caption{
Contours of the surface brightness of the Norma cluster,
as derived from ROSAT-PSPC observations, superimposed on
the {\it Swift}/BAT significance map.}
  \label{fig:norma_img}
\end{figure*}

\begin{figure*}[ht!]
  \begin{center}
  \begin{tabular}{cc}
    \includegraphics[scale=0.4]{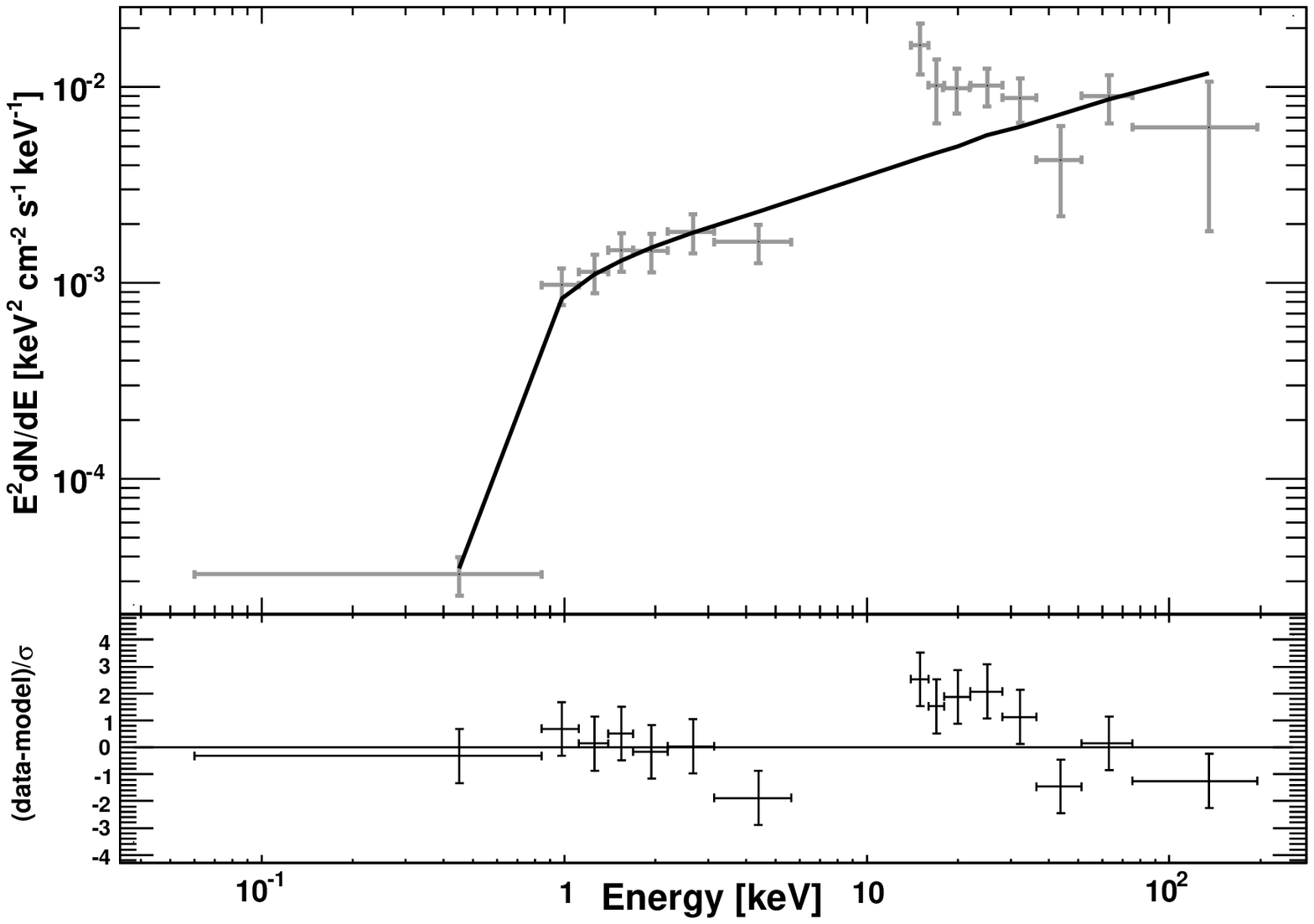}%{igr_pow.eps} 
	 \includegraphics[scale=0.4]{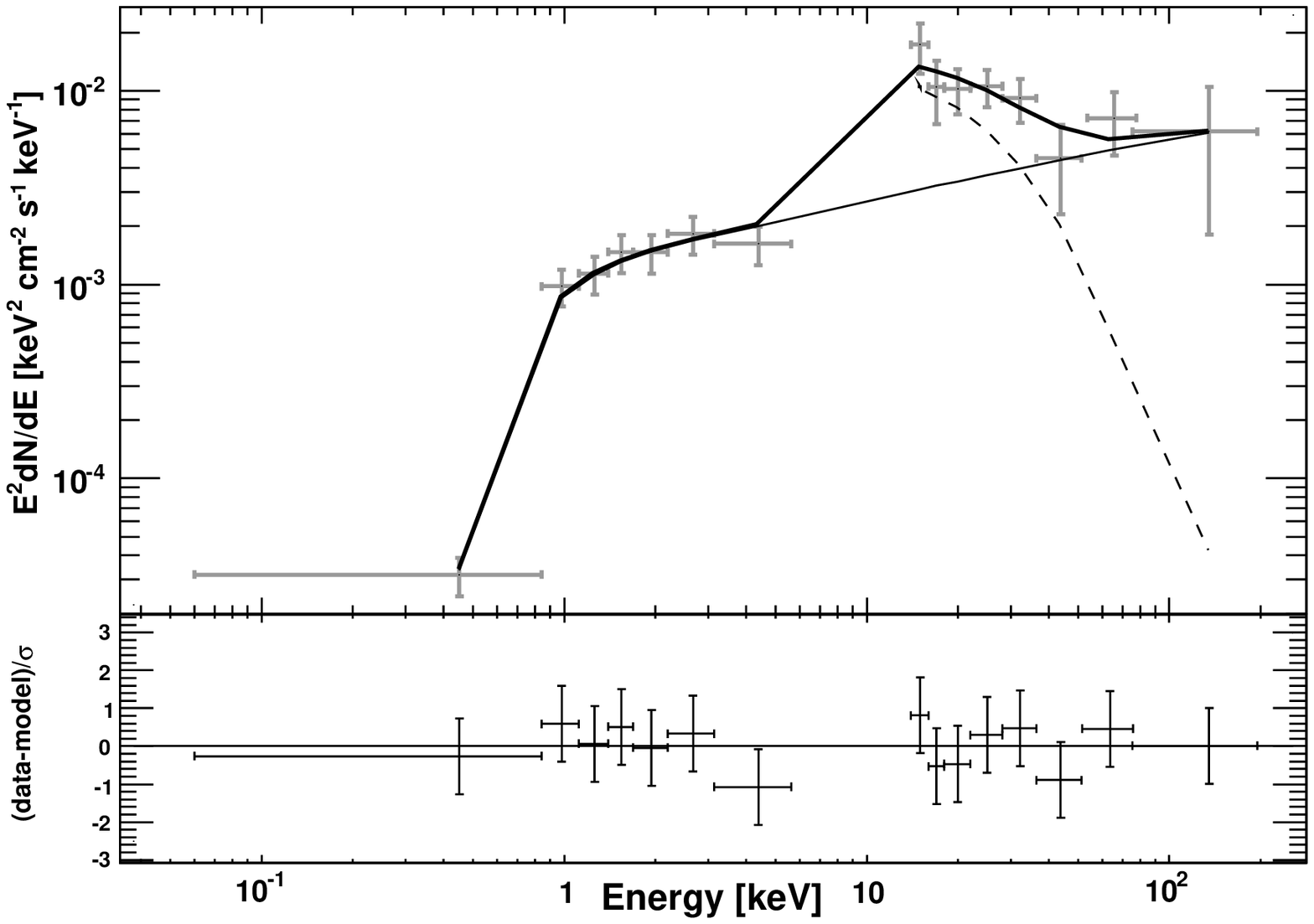}\\%{igr_brem.eps}\\
\end{tabular}
  \end{center}
  \caption{{\bf Left Panel:}
XRT and BAT spectrum of  IGR  J16119-6036 fitted with a power law.
{\bf Right Panel:}
XRT and BAT spectrum of  IGR  J16119-6036 fitted with the sum of a
 power law and a thermal model (the latter only for the BAT data).
}
  \label{fig:igr}
\end{figure*}

Considering the values for the surface brightness 
reported by \cite{boehringer96}
(e.g. core radius of 9.95\arcmin\ and $\beta=$0.55)  and Fig.~\ref{fig:ext}
it is clear that Norma should be detected, by BAT, as an extended source.
However, its elongation (e.g. not being spherical symmetric) does not
allows us to determine a-priori the expected likely flux suppression in BAT.
This is made even more complex by the presence of a nearby AGN (IGR J16119-6036) 
which is detected at a significance of $\sim10$\,$\sigma$ by BAT  
\citep{cusumano09}. This AGN is located at $\sim$20\arcmin\ away 
from the BAT centroid of Norma and the two sources appear separated.
Fig.~\ref{fig:norma_img} shows the contours of the surface brightness
of Norma (as derived from ROSAT-PSPC observations) superimposed on the 
BAT significance map for that region. It is clear from the ROSAT contours
that Norma extends likely all the way to the nearest point-like source 
(IGR J16119-6036).
We should thus expect a contamination of the cluster thermal emission
in the BAT spectrum of IGR J16119-6036. This would imply that BAT
detects the Norma cluster as an extended source.
We started fitting the BAT data alone for the Norma cluster with
a bremsstrahlung model. The fit is acceptable ($\chi^2/dof$=14.4/14)
and the best-fit temperature is  11.6$^{+6.2}_{-3.3}$\,keV.
Next, we  extracted {\it Swift}/XRT data for IGR  J16119-6036 
and fitted them together with the BAT data for this source. 
The results are reported
in Fig.~\ref{fig:igr}. When using a single absorbed power law (which
fits the XRT data alone well), the fit to the XRT-BAT dataset
 is unacceptable with $\chi^2/dof=31.5/16$ and leaves (as it can
be seen in the left panel of Fig.~\ref{fig:igr}) residuals 
in the BAT band. We then added a bremsstrahlung model (only for the BAT,
since XRT detects IGR  J16119-6036 as point-source) to the AGN power law,
to check whether the BAT spectrum is contaminated by the thermal
emission from the Norma cluster. The fit with this model is
good ($\chi^2/dof$=12.1/14) and the photon index of the power law
is 1.67$\pm0.16$ while the temperature of the thermal model is 
12.3$\pm4.2$\,keV. The temperature is in good agreement with the temperature
of the Norma cluster (as measured with BAT) reported above and thus
we conclude that there is significant contamination, particularly below 50\,keV,
of cluster's emission in the BAT spectrum of IGR  J16119-6036.
This also means that BAT detects the Norma cluster as an extended source.

\begin{figure*}[ht!]
  \begin{center}
  \begin{tabular}{cc}
    \includegraphics[scale=0.4]{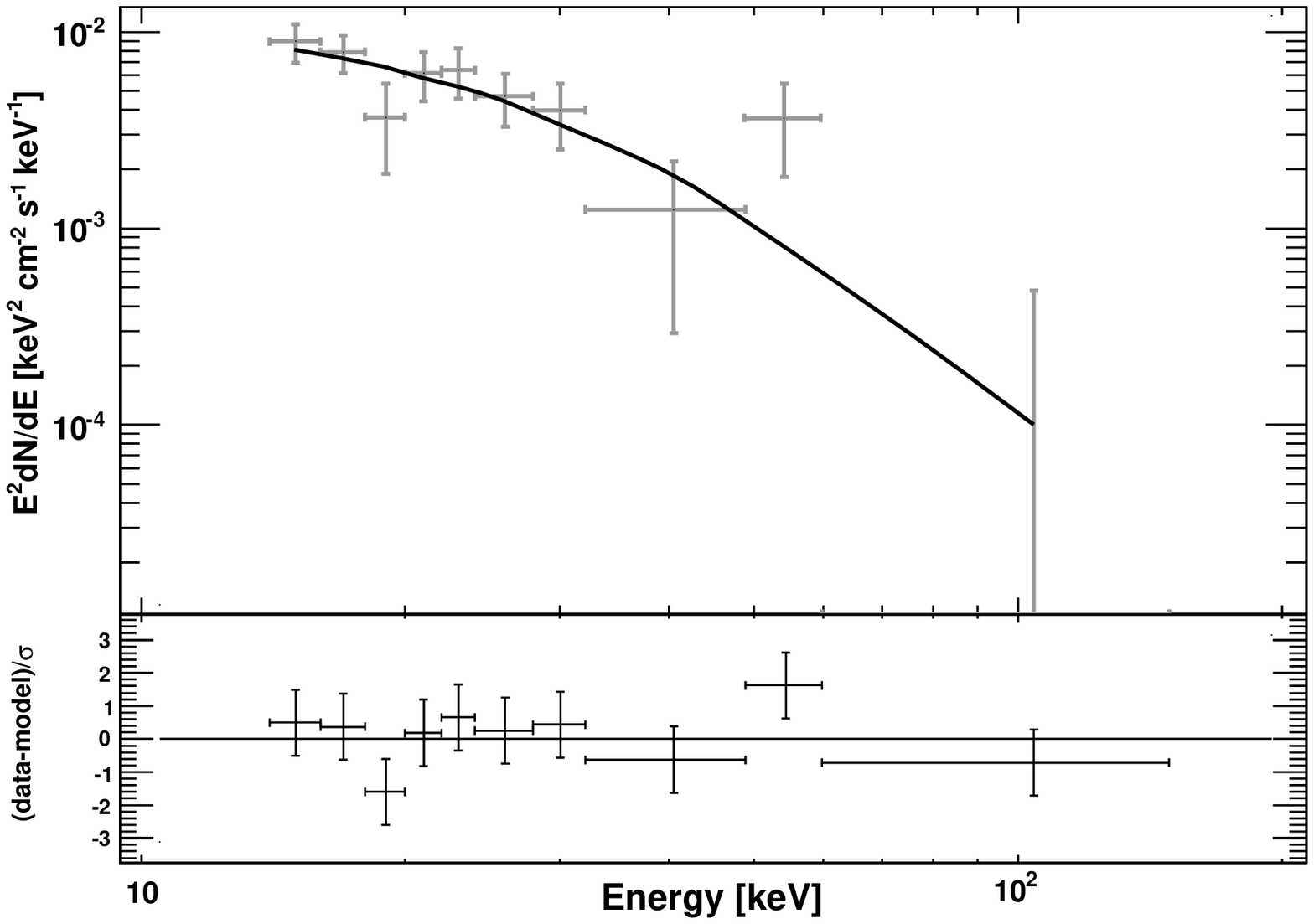}%{norma_bat.eps} 
	    \includegraphics[scale=0.4]{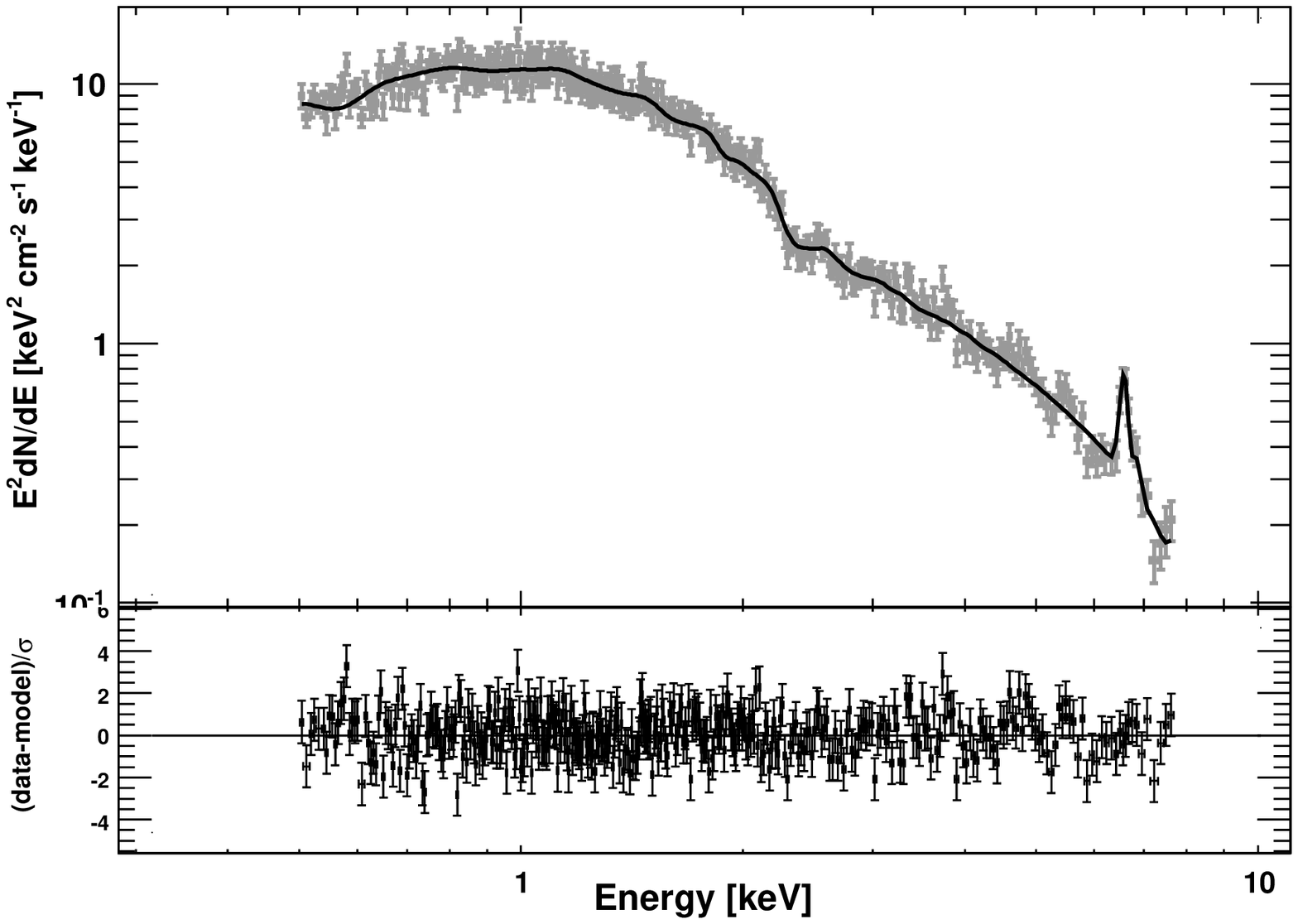}%{norma_xmm.eps}
\end{tabular}
  \end{center}
  \caption{{\bf Left Panel:} BAT data for Abell 3627 fitted
with a bremsstrahlung model.
{\bf Right Panel:}XMM-Newton data for Abell 3627 fitted with
a single thermal model.
}
  \label{fig:a3627}
\end{figure*}

Given the finding that BAT 'resolves' Norma, it becomes difficult
to determine how much flux has been suppressed by the mask and
such detailed analysis will be left to a future paper.
For this reasons, instead of providing a joint fit to XMM-Newton and BAT,
we perform two separate spectral fits.
The BAT data, as already described, are shown in the left panel of 
Fig.~\ref{fig:a3627} and are well fit with a bremsstrahlung model
with a temperature of 11.6$^{+6.2}_{-3.3}$\,keV.
The XMM-Newton data (extracted around 10\arcmin\ from the BAT centroid)
are well fit ($\chi^2/dof$=402.6/377) by an APEC model with a temperature
of 5.53$^{+0.26}_{-0.23}$\,keV and an abundance of 0.26$^{+0.06}_{-0.03}$.
This spectrum is shown in the right panel of  Fig.~\ref{fig:a3627}.
Clearly, this analysis points to a difference  in the temperature of the plasma as measured with the two instruments. This piece of 
evidence\footnote{The XMM-Newton
and BAT temperature are still compatible with each other within 
$\sim$2\,$\sigma$.} would point towards the existence of regions of hot gas
in the Norma cluster. While spatially-resolved spectroscopy is not
available for this cluster, both \cite{boehringer96} and \cite{tamura98}
find that for some regions of the cluster temperatures as high as 7--10\,keV
might exist, thus in agreement with the BAT detection.
For this cluster, we report in Tab.~\ref{tab:spec} the parameters
of the best fit to the BAT data alone.  Since it is resolved by BAT, 
part of its $>15$\,keV flux is lost in the background and an upper limit
to the non-thermal emission will not be computed.

%%%%%%%%%%%%%%%%%%%%%%%%%%%%%%%%%%%%%%%%%%%%%%%%%%%%%%%%%%%%%%%%%%%%%%%%%%
%%%%%%%%%%%%%%%%%%%%%%%%%%%%%%%%%%%%%%%%%%%%%%%%%%%%%%%%%%%%%%%%%%%%%%%%%%
\subsection{Abell 3667}

Abell 3667 is a cluster at z=0.055 discovered by HEAO \citep{pic}.  
ROSAT revealed that Abell 3667 is a
 dynamically interacting system with a significant X-ray 
emission associated with a group of galaxies which is likely merging
with the cluster. ROSAT  measured  for 
the cluster an average temperature of 
$\sim$6.5\,keV \citep{knopp}. Its dynamically complex structure has
been investigate by \cite{vikh}
using {\em Chandra}, revealing that the dense
cool core is moving with high velocity through the hotter, less
dense, surrounding gas, creating a cold front.
\citet{vikh} estimated the intensity of the magnetic field,  
in the vicinity of the shock region, to be
B$\sim$10\,$\mu$G.  The magnetic
field near the cold front is expected to be stronger and to have a very 
different structure compared to the bulk of the ICM.
These peculiar characteristics make A3667 a good candidate for the 
detection of a hard X-ray excess of non-thermal origin. This component
has, indeed, been reported in hard X-ray spectrum measured by
Beppo-SAX \citep{ff}.

Abell 3667 was recently studied in detail
by \cite{nakazawa09} using data from Suzaku. When modeling
the XIS (0.7--8.0\,keV) and the HXD (15--40\,keV) spectra they found
that a single thermal model fails to explain the hard X-ray data and
that another component is needed. This  is required
to be a very hot thermal component with T=$19.2^{+4.7}_{-4.0}$\,keV
or a power-law with an index of 1.39$^{+0.10}_{-0.17}$ \citep{nakazawa09}.

\begin{figure*}[ht!]
  \begin{center}
  \begin{tabular}{c}
    \includegraphics[scale=0.7]{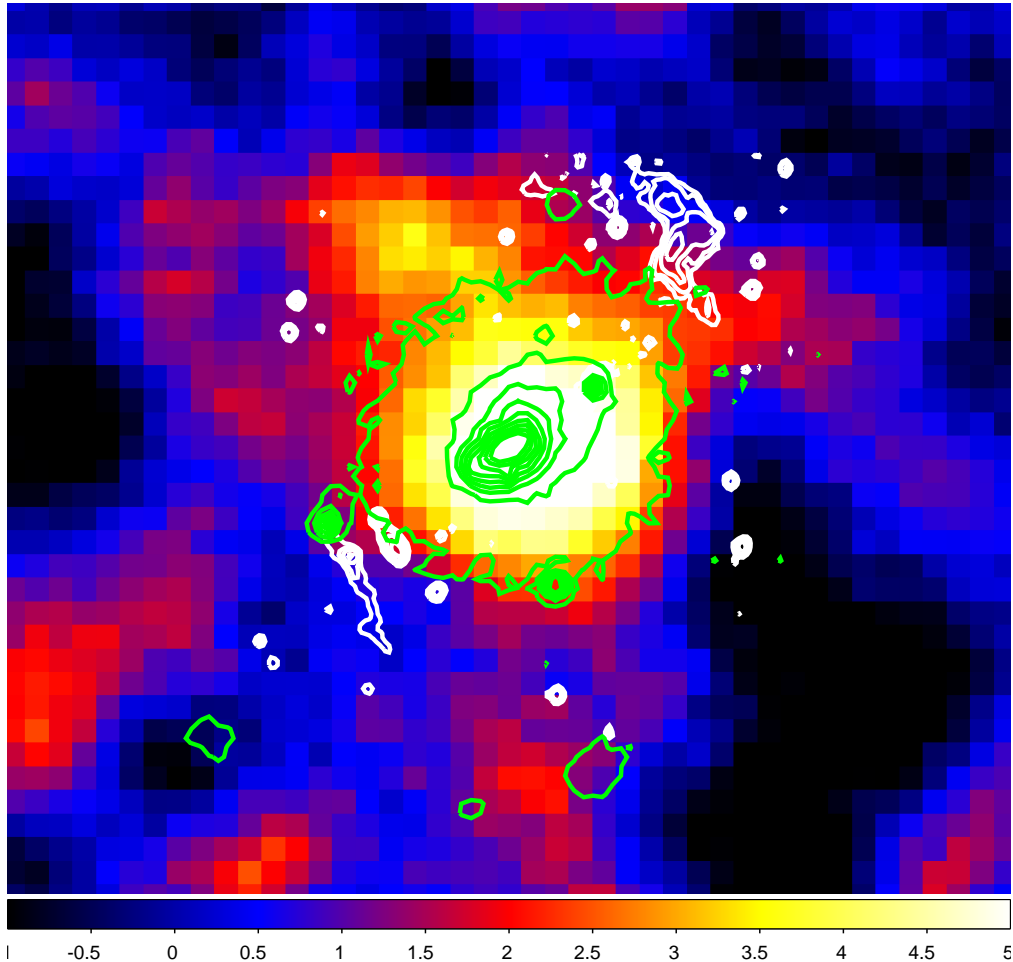}%{a3667_rosat-rad_lev.ps} 
\end{tabular}
  \end{center}
  \caption{BAT significance map of Abell 3667 with superimposed X-ray contours
from the ROSAT-PSPC (green) and radio 843\,MHz SUMSS contours (white).
}
  \label{fig:a3667_image}
\end{figure*}

\begin{figure*}[ht!]
  \begin{center}
  \begin{tabular}{c}
    \includegraphics[scale=0.7]{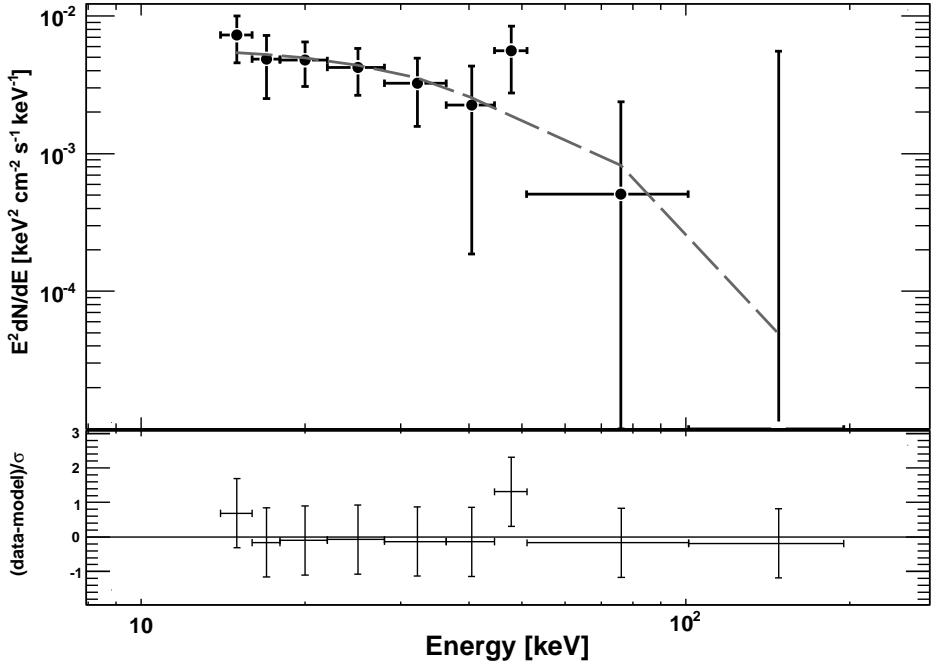} 
\end{tabular}
  \end{center}
  \caption{BAT spectrum of Abell 3667. The dashed line is the best
fit thermal model with a temperature of 18.7$^{+22.6}_{-9.3}$\,keV.
}
  \label{fig:a3667_bat}
\end{figure*}

\begin{figure*}[ht!]
  \begin{center}
  \begin{tabular}{c}
    \includegraphics[scale=0.5]{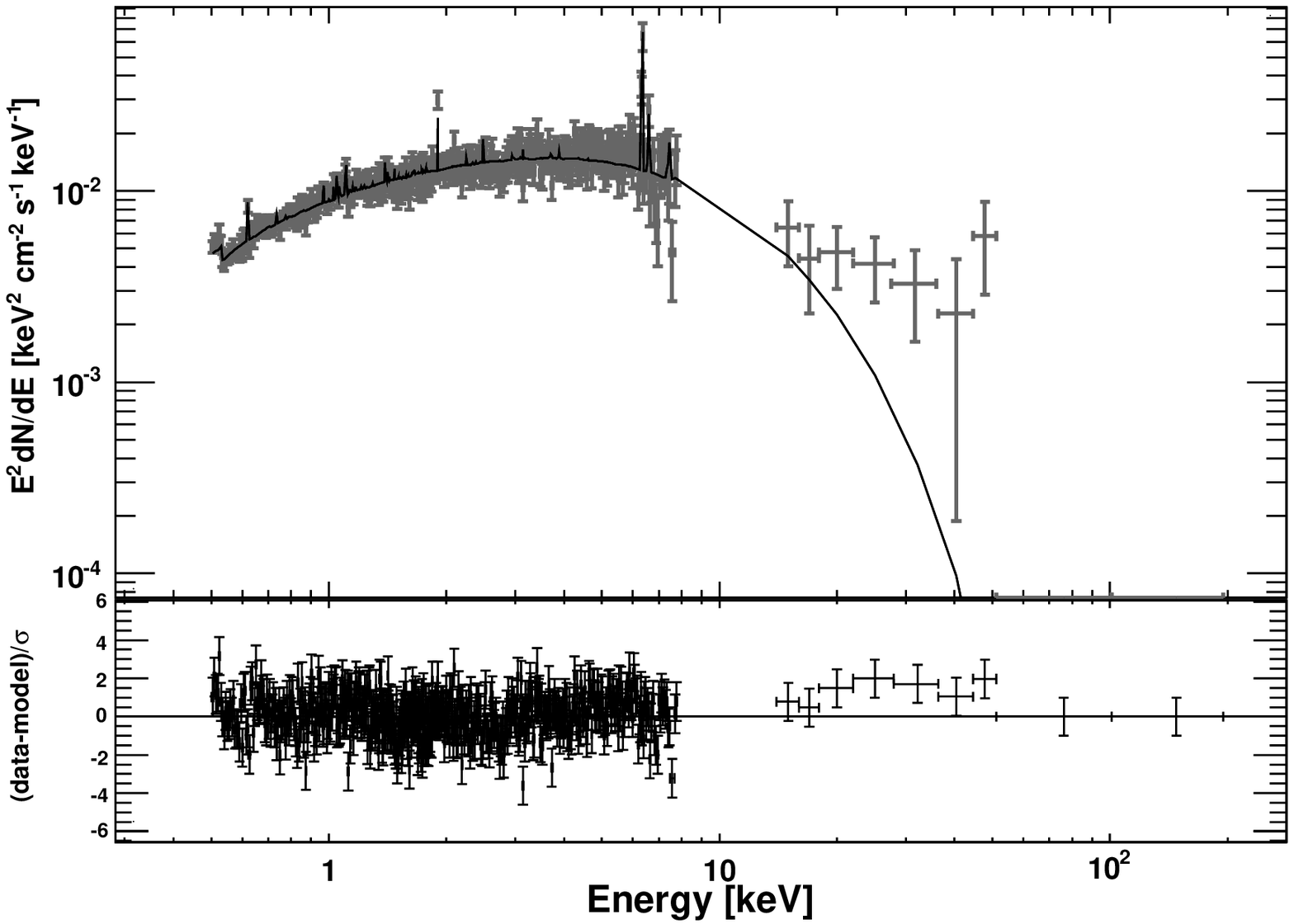}\\ %{f7a.eps} \\
	    \includegraphics[scale=0.5]{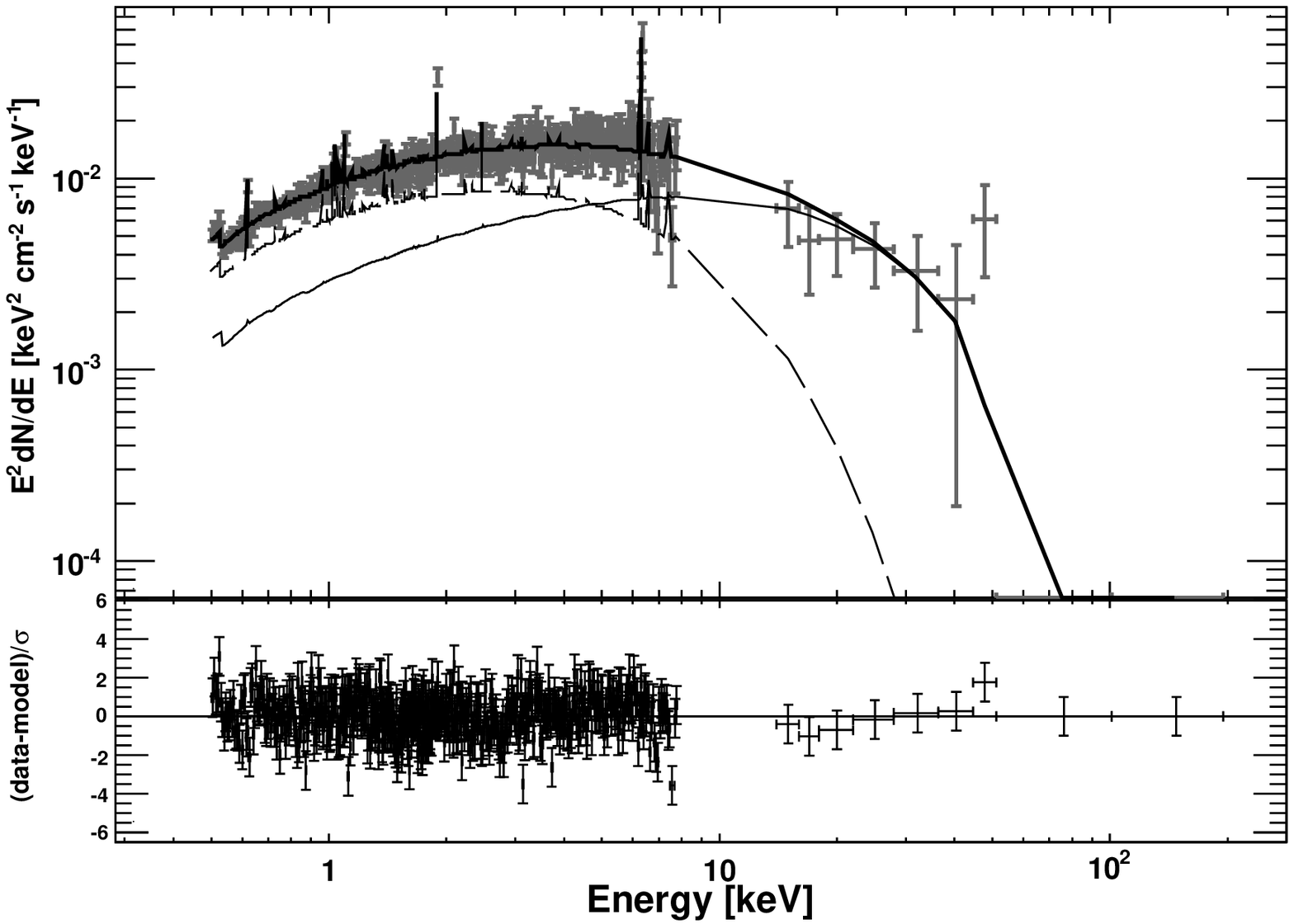}\\ %{f7b.eps} \\
 \includegraphics[scale=0.5]{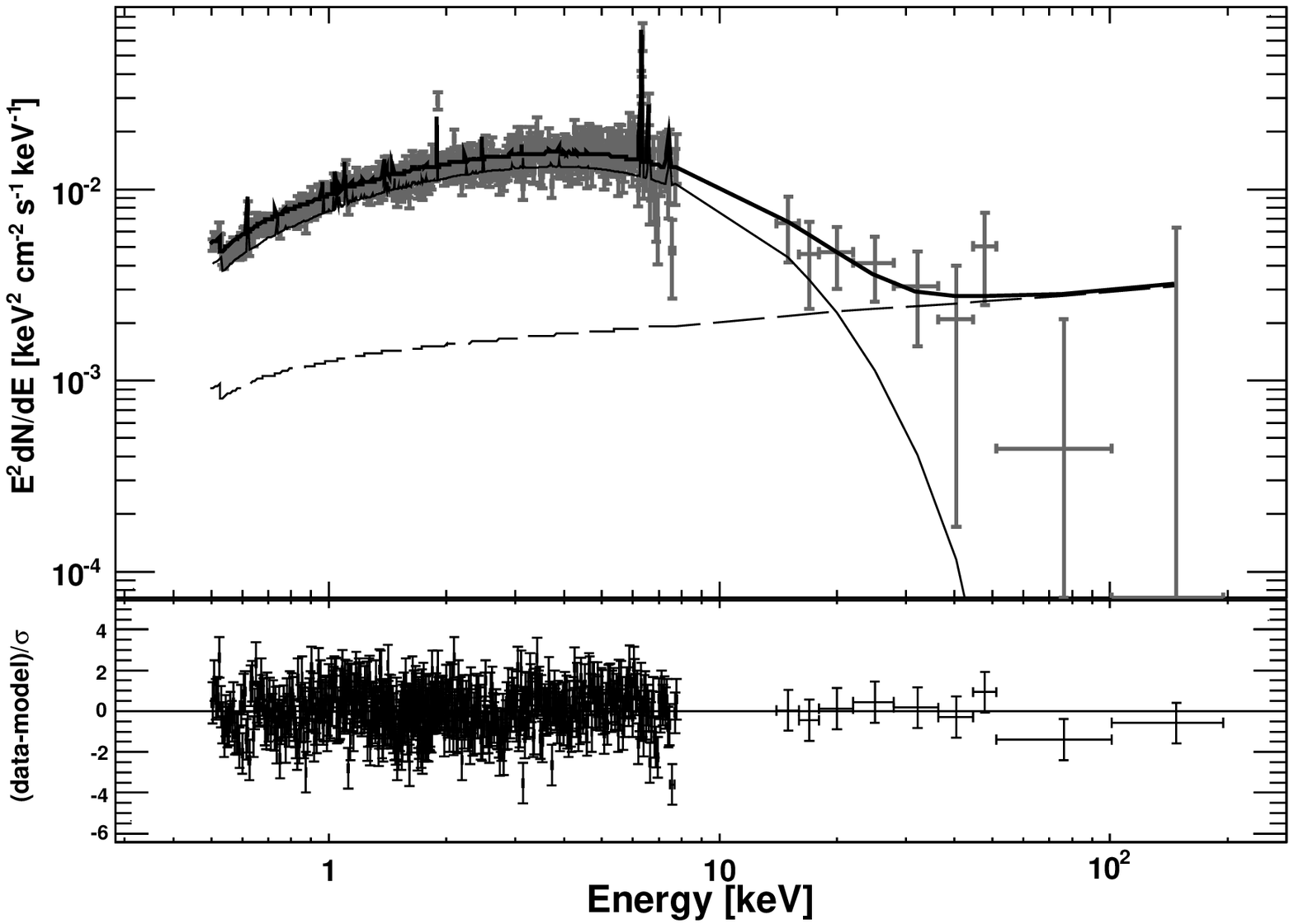}       %{f7c.eps}
\end{tabular}
  \end{center}
  \caption{XMM-Newton and BAT data for Abell 3667 fitted
with: 1) a single thermal model (top),
2) the sum of two thermal models (middle), and
3) the sum of a thermal and a power-law model.
}
  \label{fig:a3667}
\end{figure*}

Fig.~\ref{fig:a3667_image} shows the BAT significance map with superimposed
contours from ROSAT (X-rays) and SUMSS (radio). It is clear that the
BAT detection is associated with the core of the cluster and it is not
compatible as coming from the radio relic which lies $\geq$12\arcmin\
north-west of the BAT centroid.
The BAT spectrum (reported in Fig.~\ref{fig:a3667_bat})
 shows that Abell 3667 is indeed an interesting cluster.
A simple bremsstrahlung model fits the data well and
the best fit temperature is 18.7$^{+22.6}_{-9.3}$\,keV which
is unusually high even for BAT and in agreement with the one found
by Suzaku.
Adopting the values for the surface brightness reported by \cite{boehringer96}
(e.g. core radius of 2.97\arcmin\ and $\beta=$0.55) we derive that
our standard selection, in XMM-Newton, of photons within 10\arcmin\ 
includes $\sim$85--95\,\% of the cluster's emission.
When analyzing jointly XMM-Newton {\it Swift}/BAT data we find that
a single thermal model yields a best-fit temperature
of 5.68$\pm0.19$\,keV and an abundance of 0.21$\pm0.04$ solar.
This fit is reported in the upper panel of Fig.~\ref{fig:a3667}.
It is apparent that this fit leaves unsatisfactory residuals at
high-energy. We then tried adding a second thermal component.
The fit improves and the two thermal components show
a temperature of 13.5$^{+6.9}_{-2.2}$\,keV and 3.9$^{+0.8}_{-2.1}$\,keV
respectively.  The F-test shows that the probability of the second
component to be spurious is only 4.86$\times10^{-6}$.

An equally good fit can be obtained with the sum of a thermal
and a power-law model. In this case the best-fit temperature
is 5.91$\pm0.05$\,keV and the photon index is 1.83$^{+0.36}_{-0.34}$.
Again the F-test shows that the probability of the second
component to be spurious is very low, 3.46$\times10^{-6}$ (e.g.
the significance of the model is $\sim$4.6\,$\sigma$).
The power-law flux in the 10--40\,keV band is 4.91$^{+0.30}_{-2.00}\times 10^{-12}$ erg cm$^{-2}$ s$^{-1}$, a factor of $\sim$10 fainter than the one
reported by \cite{nakazawa09}.

The two models discussed here produce the same result in term of goodness
of fit and point to the existence of a very hot region with a temperature
$\sim13$\,keV (as the BAT data alone testify). On a pure statistical
basis the model with less parameters should be chosen (e.q. the thermal
plus power law model).
% and we thus
%decided to report the sum of a thermal and a power-law model in 
%Tab.~\ref{tab:spec}. 
However, on a physical basis it is difficult
to understand whether this excess is due to a 
hot component (as seen in other cases) or to a truly non-thermal
power-law like one. We believe that the hot component is the more
realistic hypothesis for several reasons: first the temperature
of this component is not unusually high for massive and merging 
galaxy clusters. Second, Abell 3667 is known to have radio relics,
but not a central radio halo \citep{rottgering97}. 
Since the BAT centroid (see Tab.~\ref{tab:xmm})
is compatible with the cluster core and not with the relic, this
renders the interpretation of the hard X-ray excess as being non-thermal
unlikely.
In support of the thermal interpretation of the hard X-ray excess we note
that the BAT centroid coincides with the hottest region (kT$\geq$8\,keV)
as found with XMM-Newton \citep{briel04}.
For this reason we decided to report the sum of the two
thermal models in Tab.~\ref{tab:spec}.

In order to exclude that this hard X-ray excess
originates from one of the point sources, we extracted
the spectrum of the brightest X-ray sources located in the cluster
field. Among all of them the brightest is 
the source positioned
 at RA(J2000)=303.14908 and Decl.(J2000)=-56.89704
with an uncertainty of 3\arcsec\ . The XMM-Newton spectrum is consistent
with a simple power-law with an index of 1.73$\pm0.20$. Its flux
extrapolated to the 10--40\,keV band is 2.68$^{+1.27}_{-0.70}\times10^{-13}$\, cm$^{-2}$ s$^{-1}$. Since this flux is a factor $>10$ fainter than the 
hard X-ray excess, we can exclude that the hard X-ray excess is caused
point-like sources. We thus conclude, partly confirming
the result of \cite{nakazawa09}, that our data requires a  hot
component (kT=13.5$^{+6.9}_{-2.2}$\,keV) or a power law with
a photon index of  1.83$^{+0.36}_{-0.34}$. In this last
case the 50--100\,keV non-thermal flux is 
2.98$^{+4.17}_{-0.73}\times10^{-12}$\,erg cm$^{-2}$ s$^{-1}$.

Finally we note that, as can be seen from Fig.~\ref{fig:a3667_image},
there is a small $\sim$3.5\,$\sigma$ fluctuation in the BAT map
 $\sim$12\arcmin\ North-East of the cluster core.
However, we remark that until the 5\,$\sigma$ threshold is exceeded
this has to be considered a statistical fluctuations.
Indeed, the probability of observing a pure $\geq$3.5\,$\sigma$ 
statistical fluctuations in the BAT map is quite large (i.e. 2.3$\times10^{-4}$)
leading\footnote{The probability of observing a fluctuation
has to be multiplied by the number of pixels in the BAT map 
\citep[i.e. 2.9$\times10^7$, ][]{segreto10}.} 
to a total of $\sim$6700 statistical fluctuations. Moreover,
no known AGN or (bright) X-ray sources are reported within 5\arcmin\ of this
fluctuation and inspection of all the available X-ray data (XMM-Newton, 
{\it Swift}/XRT, ROSAT, etc.) did not reveal any potential candidate that
might be the counterpart of this sub-threshold object.
We thus believe  this to be just a statistical fluctuation.

%%%%%%%%%%%%%%%%%%%%%%%%%%%%%%%%%%%%%%%%%%%%%%%%%%%%%%%%%%%%%%%%%%%%%%%%%%
%%%%%%%%%%%%%%%%%%%%%%%%%%%%%%%%%%%%%%%%%%%%%%%%%%%%%%%%%%%%%%%%%%%%%%%%%%
\subsection{Abell 2390}

Abell 2390 is a rich lensing galaxy cluster with a massive cool core  
\citep[e.g,]{pie96}. It is among the ten X-ray brightest galaxy 
clusters at redshift larger than 0.18 \citep[e.g,]{ebe96}. It has been 
observed with
HEAO 1 and 2 \citep{joh83,kow84,woo84,ulm86}, {\it Einstein} \citep{mcm89}, 
ROSAT \citep{pie96,ebe96,pie98,riz98,boh00},
ASCA \citep{mus97,boe98,whi00} and BeppoSAX \citep{ett01}. 
{\it Chandra} showed that Abell 2390 is experiencing
a minor merger event\citep[see e.g.][]{all01,vik05,bal07}. 
Outside the cooling region, the average temperature is 11.5\,keV  
\citep{all01}. Abell 2390 has a small (less than 2\arcmin\ ) 
irregular radio halo, most likely related to the central AGN
\citep[like the mini-halo of Perseus cluster, ][]{bac03}. 
Its flux density is 63\,mJ at 1.4\,GHz and the equipartition 
magnetic field was estimated to be 1.3\,$\mu$G \citep{bac03}.
Using the model of the surface brightness profile of Abell 2390 determined
with {\it Chandra}
\cite[e.g. see][ for details]{all01}, we derive that selecting
photons in XMM-Newton within a radius of  10\arcmin\ of the core
includes virtually all cluster's emission.

There are two bright X-ray sources located within the selection
region. Their coordinates are respectively
R.A.(J2000)=21:53:40.7  and Decl.(J2000)=17:44:13.8 for the brightest source
 R.A.=21:53:34.6 Decl.=17:36:26.8 for the dimmer one.
\cite{crawford02} conducted follow-up observations of all sources
detected with {\it Chandra} in the field of A2390.
The sources reported above correspond to the sources
A20 and A19 in their paper. Both these sources are AGN with A19 being
a Seyfert 2 galaxy at z=0.305 and A19 a QSO at z=1.6750.
The spectrum of A20  is well fit by an absorbed power law
with column density of N$_{H}=6.9^{+0.55}_{-0.30}\times10^{21}$\,cm$^{-2}$
and a photon index of 1.52$^{+0.44}_{-0.24}$. Its flux in the 2-10\,keV
band is 2.84$^{+0.80}_{-1.23}\times 10^{-13}$\,erg cm$^{2}$ s$^{-1}$ while the
extrapolated flux to the 15-55\,keV band is 5.8$\times10^{-13}$\,erg cm$^{2}$ s$^{-1}$.
The spectrum of A19 is compatible with an unabsorbed power-law model
with a photon index of 2.02$^{+0.56}_{-0.49}$. Its 2-10\,keV flux is
7.76$^{+0.45}_{-6.72}\times10^{-14}$\,erg cm$^{-2}$ s$^{-1}$ while
extrapolated flux to the 15-55\,keV band is 6.2$\times10^{-14}$\,erg cm$^{2}$ s$^{-1}$. It is clear that A19 might contribute a non-negligible fraction 
(e.g. $\sim$25\,\%) of the total flux detected in the BAT band while
that is not the case for A20. Thus when analyzing the cluster emission (below)
we will also include, in all spectral fits, an absorbed
 power-law component representing
the spectrum A19. The parameters of this absorbed power-law will be allowed
to vary within their 90\,\% CL reported above.

\begin{figure*}[ht!]
  \begin{center}
  \begin{tabular}{cc}
    \includegraphics[scale=0.4]{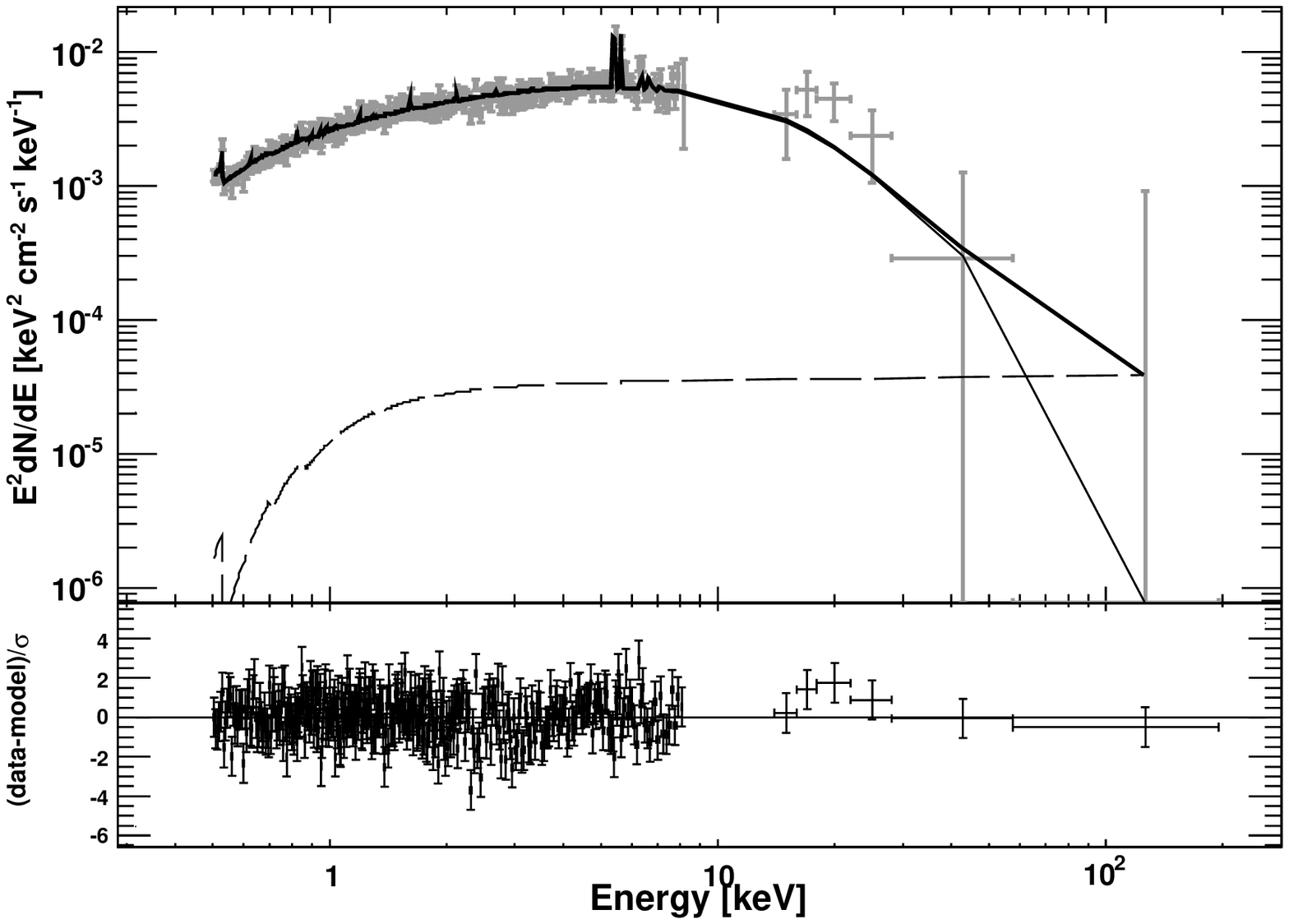}%{a2390_1T.eps} 
 \includegraphics[scale=0.4]{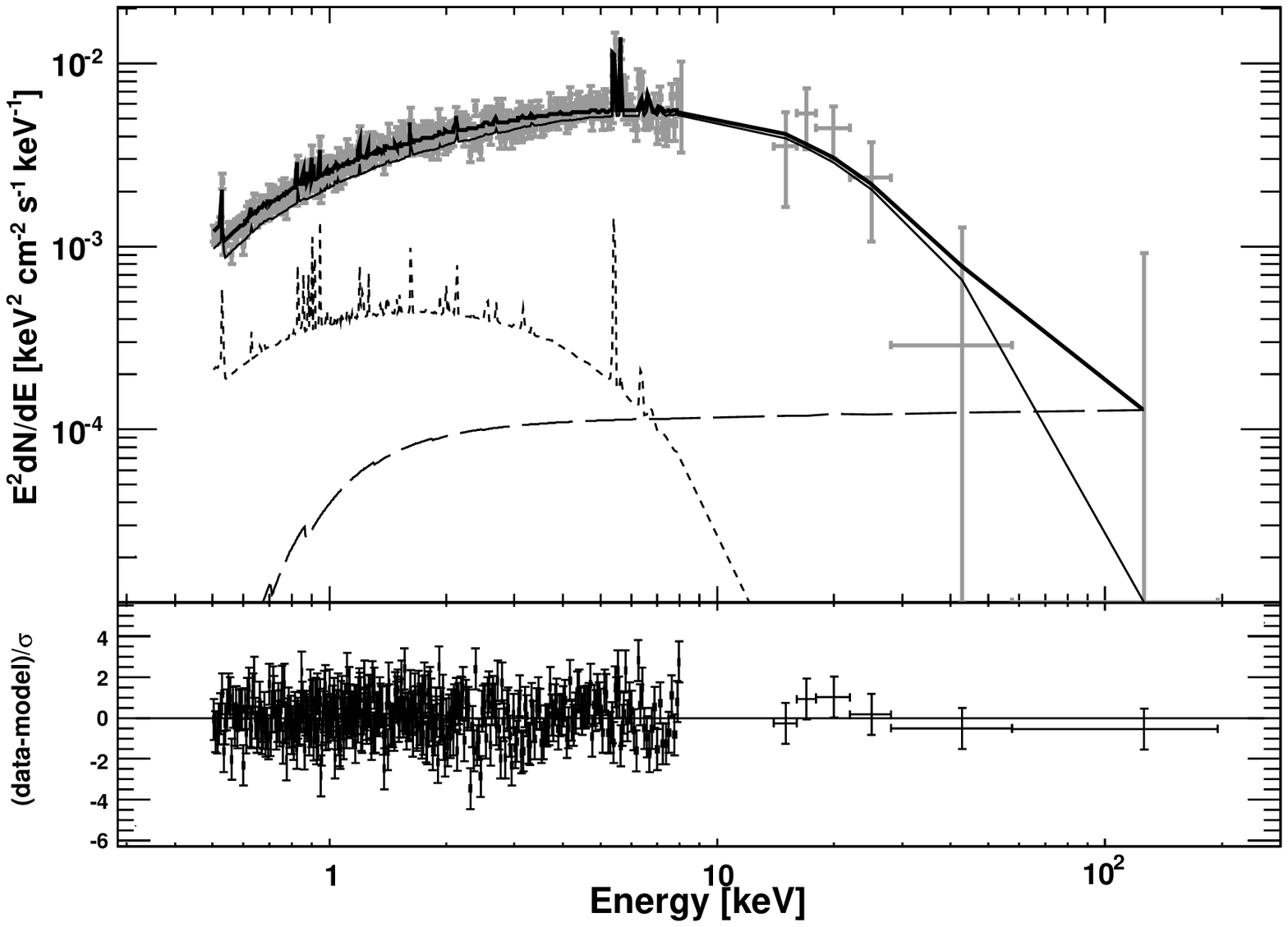}%{a2390_2T.eps} 

\end{tabular}
  \end{center}
  \caption{{\bf Left Panel:}
Spectrum of Abell 2390 fitted with a single thermal model plus
an absorbed power law (dashed line) to account for the emission
of the AGN A19 (see text for details).
{\bf Right Panel:} Spectrum of Abell 2390 fitted with
the sum of two thermal models (thin solid and short dashed
line) and an absorbed power law (dashed line) to account for the emission
of the AGN A19 (see text for details).
}
  \label{fig:a2390}
\end{figure*}

A single-temperature
plasma model  (reported in Fig.~\ref{fig:a2390}) 
with a temperature of 9.47$^{+0.43}_{-0.44}$ 
and
an abundance of 0.32$\pm0.06$ solar  successfully 
fits the XMM-Newton and {\it Swift}/BAT data. 
Our results are in good agreement with those derived 
in the 0.5--40\,keV band by BeppoSAX \citep{ett01}.
This fit is good ($\chi^2$/dof=409.9/375), but it leaves
some residuals at high energy. We then tried adding a second
thermal model and obtained a better fit (e.g. $\chi^2$/dof=394.5/372).
The improvement in the $\chi^2$ is significant and the F-test
yields a probability of $\sim$10$^{-3}$ that it was produced by chance.
The 'cold' and 'warm' components have a temperature of
 3.76$^{+2.80}_{-1.61}$\,keV and 13.08$^{+4.15}_{-2.69}$\,keV respectively,
while their abundances are 0.46$^{+0.49}_{-0.24}$ and 0.37$^{+0.23}_{-0.14}$.
Our results are in agreement with those obtained by \cite{all01} using
{\it Chandra}. Indeed, they showed that the temperature of the plasma
within 100\,kpc of the core is $\leq$5\,keV while its temperature
stays approximately constant at 11.5$^{+1.5}_{-1.6}$\,keV beyond 200\,kpc.
Adding a power law model to the baseline thermal model 
improves the fit only marginally ($\Delta \chi^2$=6.1 for
2 additional parameters), thus we consider the double-temperature thermal
model as the best representation of our dataset.
This fit, together with the single temperature thermal model, is shown in 
Fig.~\ref{fig:a2390}.

The 99\,\% CL upper limit on the 50--100\,keV  non-thermal
flux, estimated using a power-law with a photon index of 2.0,
 is 3.14$\times10^{-13}$\,erg cm$^{-2}$ s$^{-1}$.
In order to compute the lower limit on the intensity of the magnetic
field (see Tab.~\ref{tab:ulxmm}),  we adopt for this cluster
a value of the spectral index of $\alpha$=2.

\begin{deluxetable}{lccccccc}
\tablewidth{0pt}
\tabletypesize{\footnotesize}
%\rotate
\tablecaption{Spectral Properties of BAT detected Galaxy Clusters (errors are 90\% C.L.) \label{tab:spec}}
\tablehead{
%%%%%%%% column names
\colhead{NAME}                & \colhead{z}              & \colhead{Flux\tablenotemark{a}} 
& \colhead{L$_\textrm{x}$\tablenotemark{a}} & \colhead{kT} & 
\colhead{$\Gamma$/kT} & \colhead{model} 
& \colhead{$\chi^2$/dof}  \\
%%%%%%%%%  units
\colhead{}  & \colhead{}     & \colhead{\scriptsize (10$^{-12}$ cgs)}  
& \colhead{\scriptsize (10$^{43}$\,erg s$^{-1}$) }      & \colhead{ \scriptsize (keV)} 
}
\startdata
Abell 85 & 0.0521 & 5.15$^{+0.81}_{-0.83}$ & 3.81$^{+0.52}_{-0.82}$ & 6.09$^{+0.43}_{-0.29}$ & 1.72$^{+0.32}_{-0.06}$& apec + apec & 602.1/619\\

Abell  401 & 0.074 & 6.39$^{+0.91}_{-0.84}$ & 9.98$^{+1.48}_{-1.47}$ & 8.61$^{+0.60}_{-0.46}$ & 2.05$^{+0.65}_{-0.45}$ & apec + apec& 732.4/652\\  

Bullet & 0.296 & 5.10$^{+2.68}_{-1.50}$ &  176$^{+65}_{-45}$ & 14.77$^{+1.13}_{-0.72}$ & 1.86$^{+1.25}_{-0.14}$  &  apec + pow & 501.7/511 \\
%%

%%%
PKS 0745-19 & 0.103 & 6.93$^{+0.89}_{-1.16}$ & 23.2$^{+2.9}_{-3.4}$ & 7.96$^{+0.68}_{-0.54}$&  2.16$^{+1.08}_{-0.56}$ &
apec + apec & 587.1/578\\
Abell 1795 & 0.062 & 2.05$^{+0.18}_{-0.18}$ & 2.37$^{+0.23}_{-0.20}$ &  4.82$^{+0.10}_{-0.11}$ &  \nodata & apec  &  892.1/1275\\

%%%
Abell 1914 & 0.171 & 4.29$^{+1.09}_{-1.04}$ & 46.6$^{+9.9}_{-9.5}$ & 11.14$^{+1.13}_{-1.09}$& \nodata & apec & 355.1/351\\

%%%
Abell 2256 &  0.0581 & 4.46$^{+1.15}_{-1.22}$ & 4.04$^{+1.10}_{-1.07}$ & 8.84$^{+0.66}_{-0.61}$ & \nodata & apec & 445.6/445\\
%%%
Abell 3627\tablenotemark{b} & 0.0168 & 8.00$^{+1.32}_{-5.81}$ & 0.48$^{+0.08}_{-0.75}$ & 11.6$^{+6.2}_{-3.3}$ & \nodata & brem  & 14.7/14 \\
%%%

% fit con pow
%Abell 3667 & 0.0556 & 7.30$^{+1.44}_{-1.84}$ & 5.65$^{+1.07}_{-1.12}$ &  5.91$^{+0.05}_{-0.05}$ & 1.83$^{+0.36}_{-0.34}$ & apex + pow & 574.7/542 \\

% fit con 2kT
Abell 3667 & 0.0556 & 7.30$^{+1.44}_{-1.84}$ & 5.65$^{+1.07}_{-1.12}$ &  4.00$^{+0.49}_{-0.53}$ & 13.5$^{+6.9}_{-2.2}$ & apec + apec & 569.8/542 \\

%%%
Abell 2390 & 0.231 & 2.13$^{+0.26}_{-0.25}$ & 52.5$^{+5.5}_{-4.7}$ &  13.08$^{+4.15}_{-2.69}$ & 3.76$^{+2.80}_{-1.61}$& apec + apec & 394.5/372\\

\enddata

\tablenotetext{a}{Flux and Luminosities are computed in the 15--55\, keV band.}
\tablenotetext{b}{For this cluster only BAT data were used.}

\end{deluxetable}

%%%%%%%%%%%%%%%%%%%%%%%%%%%%%%%%%%%%%%%%%%%%%%%%%%%%%%%%%%%%%%%%%%%%%%%%%%%%%%%%%%
\section{Clusters Magnetic Field}
\label{sec:prop}

The diffuse synchrotron radio emission (radio halos, relics and mini-halos) proves the existence of magnetic fields
and relativistic electrons in the ICM. 
If the non-thermal X-ray emission results from IC  scattering of the same population by the CMB, then the lack of a detection of a non-thermal component can be used to place a lower limit on the magnetic fields $B$ in clusters. Indeed,
 the ratio of radio to IC flux scales proportionally to $B^{\alpha+1}$. 
Following \cite{har74},  we estimate the lower limit on $B$ (the volume averaged component along the line of sight) as explained in \cite{ajello09a}, but taking into account the redshift correction. We model the IC emission as a power law with index 2 \citep[see e.g.][for more details]{reimer04}.
The value of the   diffuse radio flux is difficult  to measure due to the presence of individual
 radio sources and to the variability of the spectral index with the distance from
  the center. Therefore the magnetic field intensities listed in Tab.~\ref{tab:ulxmm} have to be taken as order of magnitude
 estimates. We find magnetic fields that are typically a
 fraction of a $\mu$G, thus far from equipartition.  Note that A2390 is the only cluster for which we 
  evaluate the magnetic field related to the radio mini-halo--- and hence to the central AGN--- rather
   than to a more extended radio halo or radio relic.
It was not possible to estimate the lower limit on the magnetic
field intensity for a  few of the
 clusters reported in  Tab.~\ref{tab:ulxmm} for which 
there are  no detections of radio-halos reported in the literature at this time.
Given the fact that the sensitivity of BAT in its band is of the order
of $\sim 5\times 10^{-12}$\,erg cm$^{-2}$ s$^{-1}$ and thus
comparable to the sensitivities reached (in other bands) by other observatories
(e.g. HEAO-1, RXTE, Beppo-SAX etc.), 
the upper limits reported in Tab.~\ref{tab:ulxmm}
are similar to those obtained by other authors 
\citep[e.g. see][ and references therein]{rephaeli87,henriksen98,rephaeli99,ros04}.

\begin{deluxetable}{lcc}
\tablewidth{0pt}

\tablecaption{Non-thermal emission from combined XMM-Newton 
 and BAT data.
\label{tab:ulxmm}}
\tablehead{
%%%%%%%% column names
\colhead{NAME}    &
\colhead{F$_{50-100\,{\rm keV}}$\tablenotemark{a}}    & 
\colhead{B\tablenotemark{b}} \\
%%%%%%%%%  units
\colhead{}   & \colhead{\scriptsize (10$^{-12}$\,erg cm$^2$ s$^{-1}$)}  
& \colhead{\scriptsize ($\mu$G)} \\
}
\startdata
Abell 85    & $<$2.51& $\sim 0.6$ \\
Abell 401   & $<$0.22& $\sim 0.4$ \\ 
Bullet      & $$1.58$^{+0.43}_{-0.47}$& $\sim 0.16$\\
PKS 0745-19  & $<$1.6& $\sim 0.5$\\
Abell 1795  & $<$1.38& / \\
Abell 1914  & $<$1.08& $\sim 0.3$\\
Abell 2256  & $<$0.19 & $\sim 0.6$ \\
%Abell 3627  & 2.90$^{+1.78}_{-2.60}$ & / \\
Abell 3667  & 2.98$^{+4.17}_{-0.73}$ & / \\ %$\sim 0.4$ \\
Abell 2390  & $<$0.25& $\sim 0.8$\\
\enddata

\tablenotetext{a}{The flux has been estimated using a power-law spectrum
with a photon index of 2.0 in the 1--200\,keV energy band. Upper limits
are 99\,\% CL while errors are 90\,\% CL.}
%%%
%%%
\tablenotetext{b}{In order to compute the intensity of the magnetic
field we used the radio data listed in Section 2. When $\alpha$
 was not available, we adopted $\alpha=2$.}
%%%
%%%
%%%
\end{deluxetable}

%%%%%%%%%%%%%%%%%%%%%%%%%%%%%%%%%%%%%%%%%%%%%%%%%%%%%%%%%%%%%%%%%%%%%%%%%%%%%%
\section{Conclusions}

The present work combines {\it Swift}/BAT   and XMM-Newton observations  
to investigate the presence of a hard X-ray excess in the spectra of 10 
galaxy clusters detected in the ongoing BAT survey \citep{cusumano09}. Our 
results agree with our previous findings for a sample of 10 clusters  
\citep{ajello09a} ---i.e.,  most of the clusters' spectra are best
described by a multi-temperature thermal model.  
The only exception is represented by the Bullet cluster and Abell 3667, for which
 we find evidence (at the 4.4\,$\sigma$ 
and 4.6\,$\sigma$ level respectively) for a hard X-ray excess.

For the Bullet cluster, our data points to the existence of 
a power-law like component with a photon index of  1.86$^{+1.25}_{-0.14}$
and a 20--100\,keV flux of 3.4$^{+1.1}_{-1.0}\times10^{-12}$\,erg
cm$^{-2}$ s$^{-1}$. The flux of this component is found to be in good
agreement with similar values reported by \cite{pet06} and \cite{million09}.
Using the flux reported above and radio data available in the literature,
we estimate that the volume average magnetic field should have an intensity
of  $\sim$0.2\,$\mu$G.

The case of Abell 3667 is different. Indeed for  Abell 3667,
the excess can be explained in terms of a hot component with
a temperature of 13.5$^{+6.9}_{-2.2}$\,keV. Our findings
are in agreement with the results from Suzaku \citep{nakazawa09}.
The lack of a central radio halo in Abell 3667,
supports the thermal origin of the hard X-ray excess.

The Norma cluster is a special cluster for a different reason.
It is the second cluster, along with Coma \citep{ajello09a}, to be
resolved spatially by BAT. The BAT spectrum shows (albeit with low
statistics) that the temperature is around $\sim$10\,keV and thus
hotter than the temperature ($\sim$5\,keV)
determined at lower energies with XMM-Newton. Since it is resolved
by BAT a special care must be taken when analyzing data from this 
cluster and a detailed analysis will be presented elsewhere.

 Three of the detected clusters (PKS 0745-19, A1795 and A2390)  have 
 a bright cool core, while three 
 (Bullet Cluster, A2256 and A3627)   are undergoing a major merger. 
A1914 , A85 and A3667 show  signs of a minor merger as well.
Six clusters (Abell 401, Bullet Cluster, PKS 0745-19, Abell 1914, Abell  2256,
and Abell 3667)  have a radio-halo or a radio-relic.
The best spectral fits  are given by the sum of two thermal components for five
 clusters (A85, A401, PKS 0745-19,  A3667 and A2390). The other
four clusters (A1795, A1914, A2256, A2390)
 are successfully fit with a relatively hot single temperature profile.

The upper limit to the non-thermal emission (in the 50--100\,keV band)
is around $10^{-12}$\,erg cm$^{-12}$ s$^{-1}$ for most clusters.
%and this implies that the intensity of the magnetic field is a fraction of a $\mu$G.
Once again, our results indicate that the hard X-ray emission from galaxy clusters 
is mostly thermal and probably related to
post-shock regions (in the case of merging clusters) or hot regions 
outside the cool core (in the case of relaxed clusters).
It is reasonable to assume that the relativistic electrons observed 
in the radio band, produce a power-law like emission
at higher energies  due to IC scattering  off CMB photons 
\citep[e.g.][]{rep79,sar99}. In this 
case, 
%we can use the non-detection of
%non-thermal X-ray emission 
we used the presented results 
to obtain lower limits on the clusters' 
magnetic fields.  In this way, we find magnetic fields of the order 
of a fraction of $\mu$Gauss, that are far from equipartition and in 
agreement with previous similar estimates. These limits
are generally a factor 10 below the estimates obtained using 
Faraday rotation measures \citep[e.g.][]{clarke01,guidetti08,bonafede10}.

The Bullet cluster is the only one that stands out among the clusters
detected so far by BAT for the evidence of a power-law like,
 hard X-ray excess. However, many factors make this cluster special and 
unique; among them: the intermediate redshift and the violent merging
activity. The merging process powers shocks \cite[e.g.][]{mar02} 
where CRs can be accelerated efficiently. The energy density
of the CMB, whose photons constitute the targets for the electrons, 
scales with $(1+z)^4$, and thus is a factor $\sim$2.8 larger
than at redshift zero. Both things probably concur in producing the 
'bright' non-thermal component observed in this cluster.

%%%%%%%%%%%%%%%%%%%%%%%%%%%%%%%%%%%%%%%%%%%%%%%%%%%%%%%%%%%%%%%%%%%%%%%%%%%%
\acknowledgments
A critical review from the referee improved this paper drammatically.
PR is supported by the Pappalardo Postdoctoral Fellowship in Physics at MIT.
NC was partially supported from a NASA grant NNX07AV03G.
MA and PR wish to acknowledge Bal\'{u} for his positivity.
This research has made use of the NASA/IPAC extragalactic Database (NED) which
is operated by the Jet Propulsion Laboratory, of data obtained from the 
High Energy Astrophysics Science Archive Research Center (HEASARC) provided 
by NASA's Goddard Space Flight Center, and of the SIMBAD Astronomical Database
which is operated by the Centre de Donn\'ees astronomiques de Strasbourg.

{\it Facilities:} \facility{Swift (BAT/XRT)}, \facility{XMM-Newton}. 

\bibliographystyle{apj}
\bibliography{/Users/majello/Work/Papers/Clusters2/biblio}

\end{document}